\documentclass[preprint,preprintnumbers,showpacs,amsmath,amssymb]{revtex4}

\usepackage{amsfonts}
\usepackage{amssymb}
\usepackage{latexsym}
\usepackage{graphicx}% Include figure files
\usepackage{rotating}
\DeclareFontFamily{OT1}{pzc}{}
\DeclareFontShape{OT1}{pzc}{m}{it}%
              {<-> s * [1.100] pzcmi7t}{}
\DeclareMathAlphabet{\mathpzc}{OT1}{pzc}%
                                 {m}{it}
 \newcommand{\al}{\alpha}
 
 \newcommand{\be}{\beta}

 \newcommand{\si}{\sigma}
 \newcommand{\Si}{\Sigma}

 \newcommand{\de}{\delta}

 \newcommand{\De}{\Delta}

 \newcommand{\rar}{\rightarrow}

\begin{document}

\preprint{M\'exico ICN-UNAM  \, September 2009}
\title{Charged Hydrogenic, Helium and Helium-Hydrogenic Molecular Chains
in a Strong Magnetic Field}

\author{A.~V.~Turbiner}
\email{turbiner@nucleares.unam.mx}
\affiliation{Instituto de Ciencias Nucleares, Universidad Nacional
Aut\'onoma de M\'exico, Apartado Postal 70-543, 04510 M\'exico,
D.F., Mexico}
\author{N.~L.~Guevara}
\email{nicolais@nucleares.unam.mx}
\affiliation{Quantum Theory Project, Physics Dept, University of Florida,
Gainesville, FL 32611, USA}

\author{J.~C.~L\'opez Vieyra}
\email{vieyra@nucleares.unam.mx}
\affiliation{Instituto de Ciencias Nucleares, Universidad Nacional
Aut\'onoma de M\'exico, Apartado Postal 70-543, 04510 M\'exico,
D.F., Mexico}

\date{\today}

\begin{abstract}
A non-relativistic classification of charged molecular hydrogenic, helium and mixed helium-hydrogenic chains with one or two
electrons which can exist in a strong magnetic field $B \lesssim 10^{16}\,$G is given.
It is shown that for both $1e-2e$ cases at the strongest studied magnetic fields the longest hydrogenic chain
contains at most five protons indicating to the existence of the $\rm{H}_5^{4+}$ and $\rm{H}_5^{3+}$ ions, respectively. In the case of the helium chains the longest chains can exist at the strongest studied magnetic
fields with three and four $\al-$particles for $1e-2e$ cases, respectively.
For mixed helium-hydrogenic chains the number of heavy centers can reach five for highest magnetic fields studied.
In general, for a fixed magnetic field two-electron chains are more bound than one-electron ones.

\end{abstract}


% physics/0606083

\maketitle

%\mainmatter

\section{\protect\bigskip introduction}

The behavior of atoms, molecules and ions placed in a strong
magnetic field has attracted a significant attention during the last
two decades (see, for example, the review papers
\cite{Liberman:1995,Lai:2001,Turbiner:2006}). It is motivated by
both pure theoretical interest and by possible practical
applications in astrophysics and solid state physics. From the point of theory,
such studies would lead to a creation of a theory of atoms and molecules
in a magnetic field similar to a standard atomic-molecular physics. In practice,
even the basic elements of such a theory  - a knowledge of the energy levels
of the simplest Coulomb systems which can exist in a magnetic field - can be
important for interpretation of the spectra of white dwarfs (where a surface
magnetic field ranges in $B\approx 10^{6}-10^{9}$\,G) and neutron
stars where a surface magnetic field varies in $B\approx
10^{12}-10^{13}$\,G, and can be even $B\approx 10^{14}-10^{16}$\,G
for the case of magnetars.

It was conjectured long ago \cite{Kadomtsev:1971, Ruderman:1971} that unusual chemical compounds can appear in a strong magnetic field. In particular, it was suggested by M Ruderman \cite{Ruderman:1971} and then developed by his followers (see \cite{Lai:2001} and references therein) that the presence of a strong magnetic field can lead to the formation of linear hydrogenic neutral molecules (linear chains) situated along magnetic lines. It was assumed that in the ground state all electrons are in the same spin state with all spins antiparallel to the magnetic field line. To avoid a contradiction with the Pauli principle it was further assumed that all electrons have different magnetic quantum numbers. It was considered as a characteristics of the ground state. It seems obviously correct in the case of atoms and atomic ions where the electrons are close to each other. However, it is not that obvious for the case of molecules where the electrons are situated in far distant places in space. All of them (or, at least, some of them) can be in the same quantum state, with the same spin projection and magnetic quantum number \cite{Turbiner:2006London}. This situation was observed in $\rm{H}_2$ \cite{schmelcher-H2} and $\rm{H}_3^{+}$ \cite{Turbiner:2007_H3}, where in a domain of large magnetic fields the ground state was given by the state of the maximal total spin but with the electrons having the same zero magnetic quantum number (see a discussion below). In \cite{Ruderman:1971} qualitative arguments were presented that such chains can be of any length, thus, can contain arbitrary many protons. It seems that such a picture is oversimplified, it intrinsically assumes that the magnetic field is "infinitely" strong. For instance, for any exotic chain (which does not
exist in field-free case) there must be a certain threshold magnetic field since
that it begins to exists. It can well happen that such a threshold magnetic field can be beyond of realistic magnetic fields which occur in Nature. This phenomenon is absent in the qualitative theory \cite{Ruderman:1971}. Thus, some very general features of the Ruderman's picture only, like growth of the binding energies, shrinking of the size of the molecules with a magnetic field increase and maximal total electronic spin can hold for realistic high magnetic fields.

It is well known that in absence of a magnetic field, in general, the hydrogenic linear chains (polymers) do not exist with the only exception of two shortest ones, $\rm{H}_2^+$ and $\rm{H}_2$ \footnote{The $\rm{H}_3^+$-ion exists in triangular geometry}. Therefore, for each other chain there must occur a threshold magnetic field from which the chain begins to exist if it is realized. It seems natural to assume that the threshold magnetic field grows with the length of the chain which is defined by a number of heavy particles therein. At the moment, only those  $\rm{H}_2^+$ and $\rm{H}_2$ - the shortest chains - are studied in details, see e.g. \cite{Turbiner:2006} and \cite{schmelcher-H2}, respectively. The results are far more sophisticated than those predicted in a simple qualitative picture in \cite{Ruderman:1971}. For example, the $\rm{H}_2$-molecule does {\it not} exist at a large domain of strong magnetic fields.

The aim of this article is to perform a detailed quantitative study of Hydrogen, Helium and also mixed, Helium-Hydrogen linear chains with one or two electrons making an emphasis of the domain of magnetic fields $10^{2} \leq B \leq 10^7 \mbox{a.u.} (=2.35 \times 10^{16}$\ G). It is shown that in the one electron case depending on the magnetic field strength  the hydrogenic systems $\rm{H}_2^+$, $\rm{H}_3^{2+}$, $\rm{H}_4^{3+}$ and even $\rm{H}_5^{4+}$ can exist in linear geometry. It is also shown that, as the magnetic field is increased, the exotic helium-hydrogenic chains  ${\rm He}_{2}^{3+}$, ${\rm (HeH)}_{}^{2+}$, ${\rm (HHeH)}_{}^{3+}$, ${\rm (HeHHe)}_{}^{4+}$ and  ${\rm He}_{3}^{5+}$ begin to exist in linear geometry (see for a brief review \cite{Turbiner:2006London}).
For all magnetic fields the system $\rm{H}_2^+$ is stable when the system
$\rm{H}_3^{2+}$ becomes stable at $B \gtrsim 10^{13}$\,G. A detailed
review of the current status of some one-electron hydrogenic molecular systems,
both traditional and exotic, that might exist in a magnetic field $B
\geq 10^{9}$\,G can be found in \cite{Turbiner:2006}. For
two-electron case depending on the magnetic field strength the hydrogenic chains
$\rm{H}_2$, $\rm{H}_3^{+}$, $\rm{H}_4^{2+}$ and at most
$\rm{H}_5^{3+}$ can exist in linear geometry, as well as the two-electron Helium
chains ${\rm He}_2^{2+}$, ${\rm He}_3^{4+}$ and ${\rm He}_4^{6+}$ , and the mixed Hydrogen-Helium chains $({\rm HeH})^{+}$, $({\rm H-He-H})^{2+}$,$({\rm He-H-He})^{3+}$, $({\rm H-He-He-H})^{4+}$, $({\rm He-H-H-He})^{4+}$, $({\rm H-H-He-H-H})^{4+}$, $({\rm H-He-H-He-H})^{5+}$ and $({\rm He-He-H-He-He})^{7+}$. Since our study is limited to the question of the existence of a particular Coulomb system the main attention is paid to an exploration of the ground state.
Overall study is made in framework of non-relativistic consideration by solving the Schroedinger equation. It is also assumed that the Born-Oppenheimer approximation of zero order holds, which implies that the positions of positively-charged heavy particles are kept fixed (they are assumed to be infinitely-massive). Relativistic corrections are always neglected assuming that the longitudinal motion of electrons is non-relativistic for magnetic field $\lesssim 10^{16}$\,G while there are no relativistic corrections to the energies of transverse motion since the spectra of non-relativistic and relativistic harmonic oscillators coincide (we call it `the Duncan argument', for a discussion see \cite{Duncan}).
Some preliminary results were announced in \cite{Turbiner:2006London}.

Atomic units are used throughout ($\hbar$=$m_e$=$e$=1), although
energies are expressed in Rydbergs (Ry). The magnetic field $B$ is
given in a.u. with a conversion factor $B_0 = 2.35 \times 10^9$\,G.

\section{\label{hchains1e}One-electron hydrogenic chains}

\setcounter{figure}{0}

\subsection{\label{hchains1eA}Generalities}

\begin{figure}[h]
 \centering
%\fbox{
 \includegraphics[width=3.2in,angle=0]{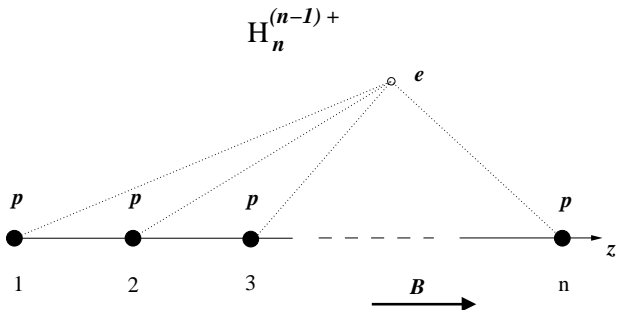}
%}
 % Hn.ps: 595x842 pixel, 72dpi, 20.99x29.70 cm, bb=0 0 595 842
 \caption{\label{figure1}${\rm H}_n^{(n-1)+}$ linear molecular ion in  parallel configuration with a magnetic field ${\mathbf B}$ oriented along the $z$-axis.}
\end{figure}

Let us consider the electron and $n$ infinitely-massive particles (protons) situated on a line which coincides to the magnetic line (see Fig.~\ref{figure1}).  We call this system a linear finite chain of the size $n$. If for such a system a bound state can be found it implies the existence of the ion $\rm{H}_n^{(n-1)+}$ in linear geometry.

%%%%%%%%%%%%%%%%%%%%  FIGURE I  %%%%%%%%%%%%%%%%%%%%%
%%%%%%%%%%%%%%%%%%%%

The Hamiltonian which describes this system when the magnetic field is oriented along the $z$ direction, ${\bf B}=(0,0,B)$ is \footnote{The Hamiltonian is
normalized by multiplying on the factor 2 in order to get the
energies in Rydbergs}
\begin{equation}
\label{H1e}
 {\cal H}_n =\left( {\hat {\mathbf p}+{\cal A}}
 \right)^2 -2 \sum_{i =1,n} \frac{Z_i}{r_{i}}\,
 + \sum_{\buildrel{i \neq j}\over {i,j =1,n}} \frac{Z_i Z_j}{R_{i j}} + 2{\bf{B}} \cdot {\bf S} \ ,
\end{equation}
(see Fig.~\ref{figure1} for the geometrical setting and notations), where $Z_i=Z_j=1$ in the case of protons, ${\hat {\mathbf p}}=-i \nabla$ is the momentum of the electron and
$\bf{S}$ is the operator of the spin, $r_{i}$ is the distance from the electron
to the $i$th proton and $R_{i j}$ is the distance between the $i$th and $j$th protons. ${\cal A}$ is a vector potential which corresponds to the constant uniform magnetic field $\bf B$. It is chosen to be in the symmetric gauge,
\begin{equation}
\label{A1e}
   {\cal A}= \frac{1}{2}({\bf{B}} \times \ {\bf{r}})\
   =\ \frac{B}{2} (-y,\ x,\ 0)\ .
\end{equation}
Finally, the Hamiltonian can be written as
\begin{widetext}
\begin{equation}
\label{H1e-final}
  {\cal H}_n = \left(- {\mathbf\nabla}^2
  +\frac{B^2}{4} \rho^2 \right) -2 \sum^n_{i =1} \frac{Z_i}{r_{i}}\,
 + \sum^n_{\buildrel{i \neq j}\over {i,j =1}} \frac{Z_i Z_j}{R_{i j}}+ B (\hat L_z +2\hat S_z )\  ,
\end{equation}
\end{widetext}
where  $\hat L_z$ and $\hat S_z$ are the z-components of the total angular
momentum and total spin operators, respectively, and $\rho=\sqrt{x^2+y^2}$. Both $\hat L_z$ and $\hat S_z$ are integrals of motion. Thus, the operators $\hat L_z$ and $\hat S_z$ in (\ref{H1e-final}) can be replaced by their eigenvalues $m$ and $m_s$ respectively. Since we are interested by the ground state for which $m=0$ and $m_s=-1/2$, the last term in (\ref{H1e-final}) can be omitted and
the reference point for energy becomes $(-B)$.

In the equilibrium  configuration the problem is characterized by two integrals of motion: (i)
angular momentum projection $m$ on the magnetic field direction
($z$-direction) and (ii) spatial parity $p$. The problem
for parallel symmetric configuration is characterized by the $z$-parity,
$P_z (z \rar -z)$ with eigenvalues $\si=\pm 1$. One can relate the magnetic quantum number $m$, spatial parity $p$ and $z$-parity $\si$,
\[
   p = \si (-1)^{|m|}\ .
\]
In the case $m$ is even, both parities coincide, $p=\si$. Thus,
any eigenstate has two definite quantum numbers: the magnetic
quantum number $m$ and the parity $p$ with respect $\vec{r} \rar
-\vec{r}$. Therefore the space of eigenstates is split into
subspaces (sectors) each of them is characterized by definite $m$
and $\si$, or $m$ and $p$. Notation for the states  is based on the
following convention: the first number
corresponds to the number of excitation - "principal quantum
number", e.g. the number 1 is assigned to the ground state, then a
Greek letter $\si, \pi, \de$ corresponds to $m=0,-1,-2$,
respectively, with subscript $g/u$ (gerade/ungerade) corresponding
positive/negative eigenvalues of spacial parity operator $P$.

\subsection{\label{hchains1eB}Method}

The variational method is used for a study of the Hamiltonian
(\ref{H1e-final}). Trial functions are chosen following the physics
relevance arguments \cite{turbinervar}. Their explicit expression is
a linear superposition of $K$ terms given by
\begin{equation}
\label{1e-psi}
 \psi_{n,K}^{(trial)} =  \sum_{k=1}^K
   A_k \Bigg\{ {e}^{-\sum^n_{i =1}  \al_{k,i} r_{i}} \Bigg\}_k
  {e}^{-  B  \be_k \frac{\scriptstyle\rho^2}{\scriptstyle 4} }\ ,
\end{equation}
(see \cite{Turbiner:2006}), where $A_k$ and $\al_{k,i}, \be_k$ are
linear and non-linear parameters, respectively. Interproton
distances $R$ are considered as variational parameters as well.
Notation $\{  \}$ means the symmetrization of identical nuclei of
the expression inside the brackets. Usually, to each term in
(\ref{1e-psi}) a certain physical meaning is given. For example, one
term had all $\al_{k,i}$ at $i=1,\ldots n$ equal being an analogue
of the Heitler-London wavefunction for the $\rm{H}_2^+$-ion -
describing the coherent interaction of the electron with all
protons. For another term all $\al_{k,i}$, except for one, vanish
being an analogue of the Hund-Mulliken wavefunction - describing the
incoherent interaction of the electron with all protons. All other
terms are different non-linear superposition of these two being an
analogue of Guillemin-Zener wavefunction for the $\rm{H}_2^+$-ion.
We call a term for which all $\al_{k,i}$ are different and unconstrained, the
{\it general term}. Needless to mention that in each particular term in (\ref{1e-psi}) the parameters are chosen in such a way to assure normalizability of this term as the overall function.

Calculations are performed using the minimization package MINUIT
from CERN-LIB. Two-dimensional integration is carried out using a
dynamical partitioning procedure: a domain of integration is
manually divided into subdomains following an integrand profile with
a localization of domains of large gradients of the integrand. Each
subdomain is integrated (for details, see, e.g., \cite{Turbiner:2006}). Numerical integration of subdomains is done with a relative accuracy of $\sim
10^{-9} - 10^{-10}$ by use of the adaptive D01FCF routine from
NAG-LIB.

\begin{enumerate}
\item[$\bf{n=1.}$]

This case was considered for the sake of completeness. It is known
that the hydrogen atom exists for any magnetic field strength. It is
the least bound system among one-electron systems. The results for
${\rm H}$-atom at $B=10^6,10^7$\,a.u. are calculated with a ten-parametric
variational trial function which is a modification of the function
introduced in \cite{turbinervar, turbiner-pot:2001}. It will be
described elsewhere.

\item[$\bf{n=2.}$]

The results for $\rm{H}_2^+$-ion are found with 3-term trial function
(\ref{1e-psi}) which depends on the 10 free parameters including the interproton distance $R$, it is a linear superposition of the Heitler-London, Hund-Mulliken and Guillemin-Zener (general term) wavefunctions. For $B\le10^4\,$ a.u. results are from \cite{Turbiner:2006}.

\item[$\bf{n=3.}$]

The results for ${\rm H}_3^{2+}$-ion are found with a 3-term trial function
(\ref{1e-psi}) which depends on  22 free parameters including two interproton distances $R$'s, it is a linear superposition of the Heitler-London,
Hund-Mulliken and a type of the Guillemin-Zener (general term) wavefunctions. For $B\le10^4\,$ a.u. results are from \cite{Turbiner:2006}.

\item[$\bf{n=4.}$]

Results for ${\rm H}_4^{3+}$-ion are found with 1-term trial function
(\ref{1e-psi}) which depends on the 7 free parameters including
three interproton distances $R$'s two of them are assumed to be equal (symmetric configuration). For $B \le 10^4\,$ a.u. the results obtained with 3- and 7-term trial function (\ref{1e-psi}) can be found in \cite{Turbiner:2006}. They lead to slightly better binding energies but do not change the qualitative picture.

\item[$\bf{n=5.}$]

It is the first study of this system. The results for the ${\rm H}_5^{4+}$-ion are obtained using a 2-term trial function (\ref{1e-psi}) which depends on the 15 free parameters including four interproton distances $R$'s, two pairs of them are assumed to be equal (symmetric configuration). In fact, it implies that a linear superposition of two general terms is taken. It is worth noting that already 1-term trial function at
$B=10^7$\,a.u. gives a clear indication to the existence of the ${\rm H}_5^{4+}$-ion with binding energy $E_b=206.11$\,Ry and equilibrium distances $R_1=0.053\,\rm{a.u.}, R_2=0.032$\,a.u. The smallest magnetic field
for which a minimum of the total energy surface in $R$'s was observed
is $5 \times 10^6$\,a.u. The ${\rm H}_5^{4+}$-ion for these magnetic fields
looks like ${\rm H}_3^{2+}$-ion bound with a far-distant proton from each side.

\item[$\bf{n=6.}$]

No indication to the existence of the ${\rm H}_6^{5+}$-ion in the domain $B \leq 10^7$\,a.u. is found.

\end{enumerate}

\subsection{\label{hchains1eC}Results}

The results of the calculations are presented in Tables ~\ref{1e_E}-\ref{1e_R}.
Two traditional for field-free case systems $\rm{H}$ and $\rm{H}_2^+$ exist
for all studied
magnetic fields $B \leq 10^7$\,a.u. The first exotic molecular
system $\rm{H}_3^{2+}$ appears at $B \sim 10^2$\,a.u. and exists for
larger magnetic fields. Another exotic molecular system
$\rm{H}_4^{3+}$ appears at $B \sim 10^4$\,a.u. and the last exotic
molecular system $\rm{H}_5^{4+}$ appears at $B \sim 5 \times
10^6$\,a.u. No other one-electron molecular hydrogenic systems are
seen for $B \leq 10^7$\,a.u. For $n>1$ the optimal geometry of any
molecular system is linear and aligned along magnetic field. Thus,
such a system forms a finite chain. It is checked that the configuration is
stable with respect to small deviations from linearity. All studied
finite chains are characterized by two features: with a magnetic
field growth (i) their total energies increase and (ii) their
lengths decrease - each system becomes more bound and compact.

For all studied magnetic fields the systems $\rm{H}$ and
$\rm{H}_2^+$ are stable: the $\rm{H}$-atom has no decay channels, where
the total energy of the $\rm{H}_2^+$-ion is always less than the total energy
 of the $\rm{H}$-atom. Furthermore, for $B \lesssim 1.5 \times 10^4$\,a.u.
the $\rm{H}_2^+$-ion has the smaller total energy then $\rm{H}_3^{2+}$-ion when exists: these two finite chains are the only ones which exists in this domain.
The $\rm{H}_3^{2+}$-ion never dissociates to $\rm{H} + 2p$ but it always
dissociates to $\rm{H}_2^+ + p$. For higher magnetic fields
$B \gtrsim 1.5 \times 10^4$\,a.u. the $\rm{H}_3^{2+}$-ion becomes stable
as well. It is characterized by the smallest total energy for these magnetic
fields. Another exotic molecular system $\rm{H}_4^{3+}$ never
dissociates to $\rm{H} + 3p$, but it dissociates to $\rm{H}_2^+ + p$
for $10^4 < B < 10^6$\,a.u. For magnetic fields $B \gtrsim
10^6$\,a.u. the total energy of $\rm{H}_4^{3+}$ is smaller than
$\rm{H}_2^+$ and the latter dissociation channel does not exist. For
all studied magnetic fields $B \leq 10^7\,$ a.u. the system
$\rm{H}_4^{3+}$ can dissociate to $\rm{H}_3^{2+}$, although the energy difference between such systems decreases gradually as the magnetic field increases.
A smooth extrapolation indicates that at the magnetic $B \sim 2\times 10^8\,$a.u. there is a crossing for which the total energies of $\rm{H}_3^{2+}$ and $\rm{H}_4^{3+}$ become equal. The system $\rm{H}_5^{4+}$ can dissociate to all finite chains except for single proton one: H-atom. Summarizing, one can state that there are two one-electron finite hydrogenic chains characterized by lowest total energy for different magnetic fields:
it is the $\rm{H}_2^+$-system at $0 \lesssim B \lesssim 1.5 \times 10^4$\,a.u.
and  the $\rm{H}_3^{2+}$-ion at $1.5 \times 10^4 \lesssim B \lesssim 10^7$\,a.u.

%%%%%%%%%%%%%%%%%%%%  TABLE I  %%%%%%%%%%%%%%%%%%%%%
\begin{table*}[ht]  % table* is used for a two column table
 \begin{center}
\small
% use packages: array
\begin{tabular}{|c||ccccccc|}  \hline
 &  &  &  &  &  &  &                \\[-7pt]
\raisebox{-5pt}{\small System} \raisebox{5pt}{\hspace{-15pt}$B$\,(a.u.)} & 0 & $1$ & $10$ & $10^{2}$ & $10^{4}$ & $10^{6}$ & $10^{7}$ \\[7pt]       \hline\hline
                       &  &  &  &  &  &       &        \\[-5pt]
 ${\rm H}_{}^{}$    & 1.0 & 1.662 & 3.495 & 7.564 & 27.10 & 73.96 & 108.86\ \\[5pt]
 ${\rm H}_{2}^{+}$  & 1.2053 & 1.9499 & 4.3498 & 10.291 & 45.799 & 139.91 & 217.75\ \\[5pt]
 ${\rm H}_{3}^{2+}$ & --      & --      & --      & 8.639  & 45.408 & 160.17 & 263.80\ \\[5pt]
 ${\rm H}_{4}^{3+}$ & --      & --      & --      & --      & 34.922 & 142.75 & 251.71\ \\[5pt]
 ${\rm H}_{5}^{4+}$ & --      & --      & --      & --      & --      & --     & 206.15\ \\[5pt]
\hline
\end{tabular}
\end{center}
\caption{\label{1e_E} Binding energies (in Ry) for the ground state $1\sigma_g$ of the one-electron hydrogenic linear systems (finite chains) in a magnetic field. Binding energies for the ground state $1s_0$ of the ${\rm H}$-atom at $0\leq B \leq 10^2\,$a.u. from \cite{turbiner-pot:2001}.}
\end{table*}

%%%%%%%%%%%%%%%%%%%%  TABLE II  %%%%%%%%%%%%%%%%%%%%%
%\begin{sidewaystable}
\begin{table*}[ht]
\small
%\begin{table}[ht]
 \begin{center}
\begin{tabular}{|c||c|c|c|c|c|c|c|}  \hline
 &  &  &  &  &  &  &                \\[-7pt]
\raisebox{-5pt}{\small System} \raisebox{5pt}{$B$\,(a.u.)} & 0 & $1$ & $10$ & $10^{2}$ & $10^{4}$ & $10^{6}$ & $10^{7}$ \\[5pt]       \hline\hline
                                 &  &  &  &  &  &  &                \\[-5pt]
 ${\rm H}_{2}^{+}\quad (R)$  &\ 1.997 &\ 1.752 &\ 0.957 &\ 0.448 &\ 0.118 &\ 0.045 &\ 0.032 \\[5pt]
 ${\rm H}_{3}^{2+}\quad (R,R)$ & --& --& --&\ 0.579, -&\ 0.130, -&\ 0.044, -  &\ 0.029, - \\[5pt]
 ${\rm H}_{4}^{3+}\quad (R_1,R_2,R_1)$ & --& --& --& --&\ 0.214, 0.138, -    &\ 0.056, 0.044 , -\,  &\ 0.034 , 0.028 , - \\[5pt]
 ${\rm H}_{5}^{4+}\quad (R_1,R_2,R_2,R_1)$ & --& --& --& --& --& --&\ 0.053, 0.032,  - , - \\[5pt]
\hline
\end{tabular}
\end{center}
\caption{\label{1e_R} Interproton equilibrium distances (in a.u.) for the ground state $1\sigma_g$ of the one-electron hydrogenic linear systems (finite chains) in a strong magnetic field. All configurations have center of symmetry, symmetric interproton distances are not displayed.}
\end{table*}
%\end{sidewaystable}

%\clearpage

\section{\label{hchains2e}Two-electron hydrogenic chains}

\subsection{\label{hchains2eA}Generalities}

\begin{figure}[h]
 \centering
%\fbox{
\includegraphics[width=3.2in,angle=0]{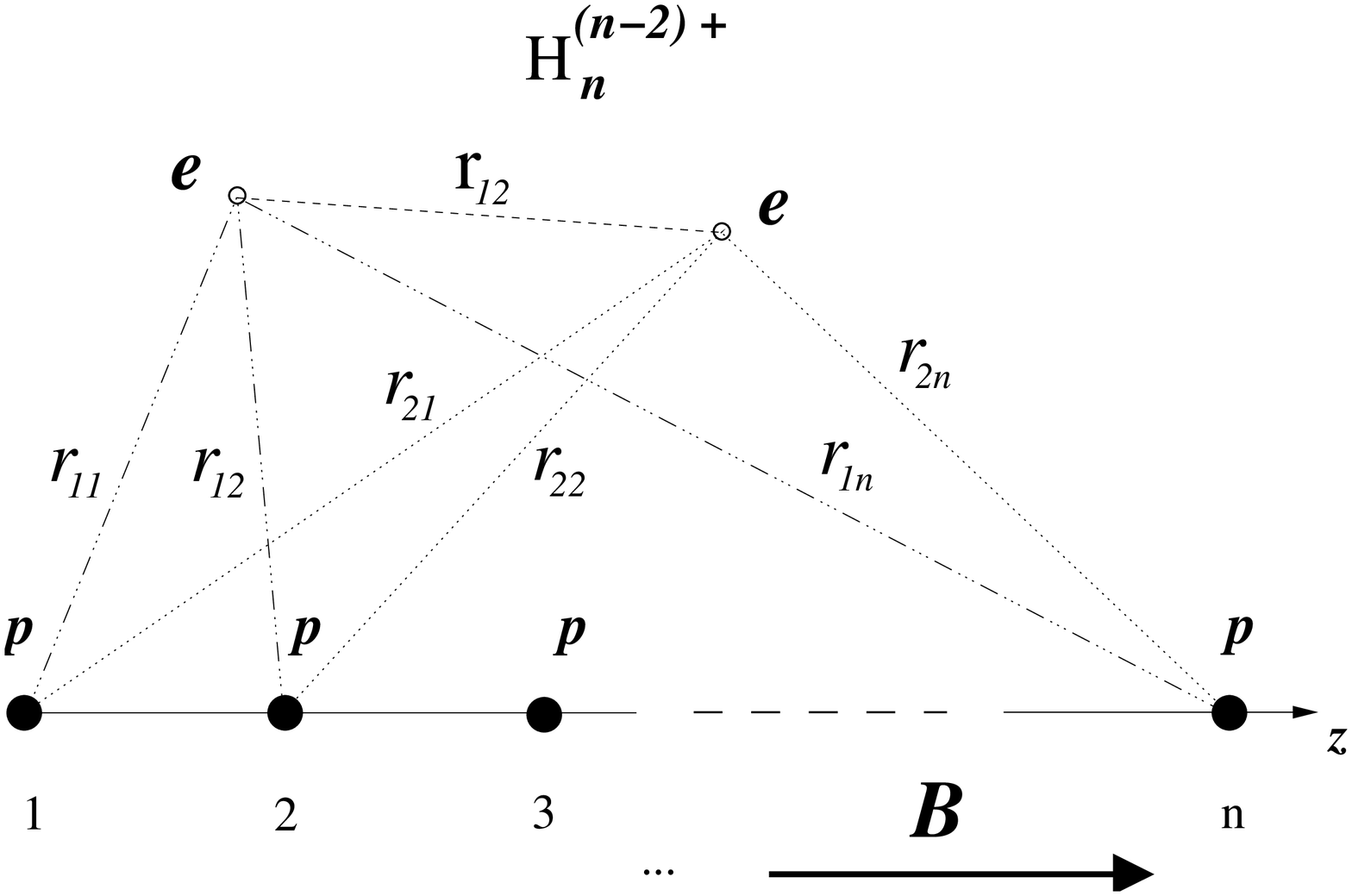}
%}
 % Hn.ps: 595x842 pixel, 72dpi, 20.99x29.70 cm, bb=0 0 595 842
 \caption{\label{figure2}${\rm H}_n^{(n-2)+}$ linear molecular ion in  parallel configuration with a magnetic field ${\mathbf B}$ oriented along the $z$-axis.}
\end{figure}

Let us consider a system of two electrons and $n$ infinitely-massive
protons situated on a line which coincides to the magnetic line (see
Fig.~\ref{figure2}).  It is called $2e$-linear finite chain of the size $n$. If
for such a system a bound state can be found it implies the
existence of the ion $\rm{H}_n^{(n-2)+}$ in linear geometry.
Sometimes, we say that above system is ``in the parallel
configuration". Also, it implies that the corresponding finite chain
exists. It can be stable or metastable.

%%%%%%%%%%%%%%%%%%%%  FIGURE II  %%%%%%%%%%%%%%%%%%%%%
%%%%%%%%%%%%%%%%%%%%

The Hamiltonian which describes the system of two electrons and $n$
protons when the magnetic field is oriented along the $z$ direction,
${\bf B}=(0,0,B)$ is~[19]

%
% \begin{widetext}
% \begin{equation}
% \label{H2e}
%  {\cal H} =\sum_{\ell=1}^2 \left( {\hat {\mathbf p}_{\ell}+{\cal A}_{\ell}}
%  \right)^2 - 2 \sum_{\buildrel{{\ell}=1,2}\over{\buildrel{i =1,n}\over{\scriptscriptstyle \hspace{7pt}=a,b\ldots}}}
%  \frac{ Z_i}{r_{{\ell}\, i}}+ \frac{2}{\mathpzc{r}_{12}} \,
%  + \sum_{\buildrel{i \neq j}\over {\buildrel{i,j =1,n}\over{\scriptscriptstyle \hspace{10pt}=a,b\ldots}}} \frac{Z_i Z_j}{R_{i,j}}  + 2{\bf{B}} \cdot {\bf S} \ ,
% \end{equation}
% \end{widetext}

\begin{widetext}
 \begin{equation}
 \label{H2e}
  {\cal H}_n =\sum_{\ell=1}^2 \left( {\hat {\mathbf p}_{\ell}+{\cal A}_{\ell}}
  \right)^2 - 2 \sum_{\buildrel{{\ell}=1,2}\over{i =1,n}}
  \frac{ Z_i}{r_{{\ell}\, i}}+ \frac{2}{\mathrm{r_{12}}} \,
  + \sum_{\buildrel{i \neq j}\over {i,j =1,n}} \frac{Z_i Z_j}{R_{ij}}  + 2{\bf{B}} \cdot {\bf S} \ ,
 \end{equation}
\end{widetext}
(see Fig.~\ref{figure2} for the geometrical setting and notations),
where $Z_i=Z_j=1$ in the case of protons, ${\hat {\mathbf p}_{\ell}}=-i
\nabla_{\ell}$ is the  momentum of the ${\ell}$th
electron, $r_{\ell i}$ is the distance from the $\ell$th electron
to the $i$th proton and $R_{i j}$ is the distance between $i$th and
$j$th proton, $\mathrm{r_{12}}=|\vec{r_1}-\vec{r_2}|$ is the interelectron
distance, where $\vec{r_{1}}\ (\vec{r_{2}})$ is the position from
the center of the chain (mid-point with respect to the end-situated
protons) of the first (second) electron and
$\mathbf{S}= \mathbf{S}_{1}+ \mathbf{S}_{2}$ is the operator
of the total spin. ${\cal A}_{\ell}$ is a
vector potential which corresponds to the constant uniform magnetic
field $\bf B$ written in the symmetric gauge (\ref{A1e}).
Finally, the Hamiltonian can be written as
\begin{widetext}
\begin{equation}
\label{H2e-final}
 {\cal H}_n =\sum_{{\ell}=1}^2 \left(- {\mathbf\nabla}^2_ {\ell}
  +\frac{B^2}{4} \rho_{\ell}^2 \right)
  - 2 \sum_{\buildrel{{\ell}=1,2}\over{i =1,n}}
 \frac{Z_i}{r_{{\ell}\, i}}+ \frac{2}{\mathrm{r_{12}}} \,
 + \sum_{\buildrel{i \neq j}\over {i,j =1,n}} \frac{Z_i Z_j}{R_{i,j}}
 + B (\hat L_z +2\hat S_z )\  ,
\end{equation}
\end{widetext}
where  $\hat L_z=\hat L_{z_1}+\hat L_{z_2}$ and $\hat S_z=\hat
S_{z_1}+\hat S_{z_2}$ are the z-components of the total angular
momentum and total spin, respectively, and
$\rho_{\ell}=\sqrt{x_{\ell}^2+y_{\ell}^2}$.
All performed calculations released a symmetry property of a chain: in the optimal geometry a chain has a center of symmetry. Hence, for any proton there is a
partner situated symmetrically with respect to this center. We consider that
property as intrinsic of any chain.

The problem under study is characterized by three conserved
quantities: (i) the operator of the $z$-component of the total
angular momentum (projection of the angular momentum on the magnetic
field direction) giving rise to the magnetic quantum number $m$,
(ii) the spatial parity operator $P({\vec r_1} \rar -{\vec
r_1},{\vec r_2} \rar -{\vec r_2})$ which has eigenvalues $p=\pm
1$(gerade/ungerade)
%(ii) the square of the total spin operator giving rise to the spin
%quantum number $S$ and
(iii) the operator of the $z$-component of the total spin
(projection of the total spin on the magnetic field direction)
giving rise to the total spin projection $m_s$. Hence, any
eigenstate has three explicit quantum numbers assigned: the magnetic
quantum number $m$, the total spin projection $m_s$ and the parity
$p$. For the case of two electrons the total spin projection $m_s$
takes values $0,\pm 1$.

As a magnetic field increases a contribution from the Zeeman term
(interaction of spin with magnetic field, ${\bf{B}} \cdot {\bf{S}}$)
becomes more and more important. It seems natural to assume that for
small magnetic fields a spin-singlet state is the state of  lowest
total energy, while for larger magnetic fields it should be a
spin-triplet state with $m_s=-1$, where the electron spins are
antiparallel to the magnetic field direction ${\bf{B}}$. The total
space of eigenstates is split into subspaces (sectors), each of them
is characterized by definite values of $m$, $p$ and $m_s$.  It is
worth noting that the Hamiltonian ${\cal H}_n$ is invariant with respect
to reflections $P_z$: $z_1 \to -z_1$ and $z_2 \to -z_2$, with
eigenvalues $\si_N=\pm 1$, for a symmetric chain.

In order to classify eigenstates we follow the convention widely
accepted in molecular physics using the quantum numbers $m, p$ and
the total spin $S$ without indication to the value of $m_s$.
Eventually, the notation is ${}^{2S+1} M_{p}$, where $2S+1$ is the
spin multiplicity which is equal to $1$ for spin-singlet state
($S=0$) and $3$ for spin-triplet ($S=1$), as for the label $M$ we
use Greek letters $\Si, \Pi, \De$ that mark the states with $|m|=0,
1, 2,...$, respectively, but implying that $m$ takes negative values
and the subscript $p$ (the spatial parity quantum number) takes
gerade/ungerade($g/u$) labels describing
positive $p=+1$ and negative $p=-1$ parity, respectively. There
exists a relation between the quantum numbers corresponding to the
$z$-parity and the spatial parity:
\[
 p\ =\ (-1)^{|m|}\ \si_N \, .
\]
Present consideration is limited to the states with magnetic
quantum numbers $m=0,-1$ because the total energy of the lowest
energy state in any sector with positive $m>0$ is always
larger than one with  $m \leq 0$. A study of states with different
$m$ is necessary to identify the state of lowest total energy. At
large magnetic fields for all studied two-electron chains this state was
characterized by $m=-1$ in agreement with Ruderman's hypothesis.

\subsection{\label{hchains2eB}Method}

As a method to explore the problem we use the variational procedure.
The recipe of choice of trial functions is based on physical
arguments \cite{turbinervar}. As a result the trial function for the
lowest energy state with magnetic quantum number $m$ is chosen in
the form
\begin{widetext}
\begin{equation}
\label{psi}
 \psi^{(trial)} =
       (1+\si_e P_{12})\,
%       (1+\si_{\scriptscriptstyle N} P_{1,n})
%       (1+\si_{\scriptscriptstyle N_a} P_{AB}
%         +\si_{\scriptscriptstyle N_a} P_{BC})\times
\rho_1^{\mid m \mid}e^{im\phi_1} \,\,  \sum_{k=1}^K
   A_k \Bigg\{
     {e}^{-\sum_{\buildrel{{\ell}=1,2}\over{i =1,n}}
     {\scriptstyle \al_{k, {\ell}\, i}} {r_{{\ell}\, i}}}
       \Bigg\}_k
   {e}^{ \gamma_k \mathrm{r_{12}} -  B  \be_{k,1}\frac{\scriptstyle \rho_1^2}{\scriptstyle 4}-
  B \be_{k,2}\frac{\scriptstyle\rho_2^2}{\scriptstyle 4}}
\end{equation}
% \begin{eqnarray}
% \label{psi}
%  \psi^{(trial)} &=&
%        (1+\si_e P_{12})
% %       (1+\si_{\scriptscriptstyle N} P_{1,n})
% %       (1+\si_{\scriptscriptstyle N_a} P_{AB}
% %         +\si_{\scriptscriptstyle N_a} P_{BC})\times
%  \non
%  \\
%  \hspace{20pt}\rho_1^{\mid m \mid}e^{im\phi_1} \,\,  &&\hspace{-22pt}\sum_{k=1}^K
%    A_k \Bigg\{
%      {e}^{-\sum_{\buildrel{{\ell}=1,2}\over{i =1,n}}
%      {\al_{k, {\ell}\, i}} {r_{{\ell}\, i}}}
%        \Bigg\}_k
%    {e}^{ \gamma_k r_{12} -  B  \be_{k, 1}\frac{\rho_1^2}{4}-
%   B \be_{k, 2}\frac{\rho_2^2}{4}}
% \end{eqnarray}
\end{widetext}
where $\si_e=\pm 1$ stands for spin singlet (+) and triplet states
$(-)$, while $\{  \}$ means the symmetrization of identical nuclei
of the expression inside the brackets. The $P_{12}$ is the
permutation operator for the electrons, (1 $\leftrightarrow$ 2). The
$\al_{k, ij}$, $\be_{k, 1-2}$ and $ \gamma_k$ as well as interproton
distances $R_{ij}=R_{ji}$ are variational parameters. For each term
with fixed $k$ their total number is $2n+4$ including the linear
parameter $A_k$. In addition, we have $n-1$ interproton distances.
It is worth emphasizing that in the trial function (\ref{psi}) the
interelectron interaction is included explicitly in the exponential
form $e^{\gamma \mathrm{r_{12}}}$.

Calculations are performed using the minimization package MINUIT
from CERN-LIB. Multidimensional integration is carried out using a
dynamical partitioning procedure: a domain of integration is
manually divided into subdomains following an integrand profile with
a localization of domains of large gradients of the integrand. Each
subdomain is integrated separately using parallelization procedure
(for details, see, e.g., \cite{Turbiner:2006}). Numerical
integration of subdomains is done with a relative accuracy of $\sim
10^{-6} - 10^{-7}$ by use of the adaptive D01FCF routine from
NAG-LIB. A process of minimization for each given magnetic field and
for any particular state was quite time-consuming due to a
complicated profile of the total energy surface in the parameter
space but when a minimum is found it takes several seconds of CPU time
to compute a variational energy.

%%%%%%%%%%%%%%%%%%%%  TABLE III  %%%%%%%%%%%%%%%%%%%%%
\begin{table*}[ht]
 \begin{center}
\small
% use packages: array
\begin{tabular}{|c||cccccc|}  \hline
 &  &  &  &  &  &         \\[-7pt]
\raisebox{-5pt}{\small System} \raisebox{5pt}{\hspace{-15pt}$B$\,(a.u.)}
 & $10^{2}$ & $10^{3}$ & $10^{4}$ & $4.414\times 10^{13}$\,G & $10^{6}$
 & $10^{7}$ \\[7pt]
\hline\hline
                    &  &  &  &  &  &                \\[-5pt]
 ${\rm H}_{}^{-}$   & 8.35   & 16.95 & 30.1 & 35.4 & 82.5 \    &\ 121.4 \\[5pt]
 ${\rm H}_{2}^{}$   & 16.473${}^{s}$ & 35.632 & 71.42 & 85.00\ & 219.9 &\ 330.3 \\[5pt]
 ${\rm H}_{3}^{+}$  & 18.915 & 44.538 & 95.21 & 115.19 & 324.2\ &\ 529.8 \\[5pt]
 ${\rm H}_{4}^{2+}$ & (17.601) & (43.917) & 99.80 & 122.34 & 367.7\ &\ 636.0 \\[5pt]
 ${\rm H}_{5}^{3+}$ & -- & -- & 91.70 & 114.34 & 383.2\ &\ 687.7 \\[5pt]
\hline
\end{tabular}
\end{center}
\caption{\label{2e_E} Double ionization energies $E_I$ in Ry ($E_T =
- E_I$) for the ground state ${}^3\Pi_u$ of the two-electron
hydrogenic systems (finite chains) in a strong magnetic field.
${}^{s}$ from \cite{schmelcher-H2}. Energy in brackets means that
the state ${}^3\Pi_u$ is bound but the ground state corresponds to
an unbound state. The magnetic field $B_{Schwinger}=4.414\times 10^{13}\,{\rm G} = 1.878 \times 10^{4} \,{\rm a.u.}$ corresponds to the so called {\it non-relativistic} threshold for which the electron cyclotron energy equals the electron rest mass.}
\end{table*}

%%%%%%%%%%%%%%%%%%%%  TABLE IV  %%%%%%%%%%%%%%%%%%%%%
\begin{sidewaystable}
%\begin{table*}[ht]
\small
 \begin{center}
% use packages: array
\begin{tabular}{|c||c|c|c|c|c|c|}  \hline
 &  &  &  &  & &                \\[-7pt]
\raisebox{-5pt}{\small System} \raisebox{5pt}{$B$\,(a.u.)} & $10^{2}$ & $10^{3}$ & $10^{4}$ & $4.414\times 10^{13}$\,G & $10^{6}$ & $10^{7}$ \\[5pt]       \hline\hline
                              &  &  &  &  & &               \\[-5pt]
 ${\rm H}_{2}^{}\quad (R)$  & 0.38${}^{s}$  & 0.19 & 0.102 & 0.087 & 0.038 & 0.034 \\[5pt]
 ${\rm H}_{3}^{+}\quad (R,R)$ & 0.395, - & 0.183, - & 0.093, - & 0.078, - & 0.030, -  &  0.023, - \\[5pt]
 ${\rm H}_{4}^{2+}\quad (R_1,R_2,R_1)$ &(0.51, 0.38, -) & (0.215, 0.174, -) & 0.103, 0.086, - & 0.092, 0.075, - &   0.030, 0.018 , - & 0.020, 0.013, - \\[5pt]
 ${\rm H}_{5}^{3+}\quad (R_1,R_2,R_2,R_1)$ & -- & -- & 0.184,0.134 , - , - & 0.160,0.110 , - , - & 0.035,0.025,-,- & 0.023 , 0.018 , - , -\\[5pt]
\hline
\end{tabular}
\end{center}
\caption{\label{2e_R} Interproton equilibrium distances (in a.u.) for the ground state ${}^3\Pi_u$ of the two-electron hydrogenic linear systems (finite chains) in a strong magnetic field. All configurations have center of symmetry and symmetric interproton distances are not displayed. ${}^{s}$ from \cite{schmelcher-H2}. Distances in brackets mean that
the state ${}^3\Pi_u$ is bound but the ground state corresponds to
 an unbound state.}
%\end{table*}
\end{sidewaystable}

\begin{enumerate}

\item[$\bf{n=1.}$]

This case corresponds to the negative hydrogen ion $\rm{H}^-$ and is mentioned for the sake of completeness. It is known that the negative hydrogen ion $\rm{H}^-$ exists for any magnetic field strength \cite{schmelcher-H-}. At zero and small magnetic fields $B < 5 \times 10^{-2}$\ a.u. the spin-singlet state ${}^1 0$
is the ground state. If $B > 5 \times 10^{-2}$ a.u. the spin-triplet
state ${}^3(-1)$ which does not exist in the absence of a magnetic
field becomes bound and the ground state. Although this result is checked
quantitatively for magnetic fields up to 4000\,a.u.
\cite{schmelcher-H-, Turbiner:unpublished} it is quite likely that
it holds for higher magnetic fields. It is the least bound system
among two-electron systems made from protons. However, the
$\rm{H}^-$-ion is stable for studied magnetic fields: the
dissociation $\rm{H}^- \rightarrow \rm{H} + e$ is prohibited.

\item[$\bf{n=2.}$]

In a domain of non ultra-high magnetic fields the
$\rm{H}_2$-molecule was studied in details in \cite{schmelcher-H2}.
It was shown that the lowest total energy state depends on the
magnetic field strength. It evolves from the spin-singlet
${}^1\Si_g$ state at $0 \leq B \lesssim 0.18$\,a.u. to a repulsive
spin-triplet ${}^3\Si_u$ state (unbound state) for
$0.18\,\mbox{a.u.} \lesssim \ B \lesssim 12.3$\,a.u. and, finally,
to a strongly bound spin-triplet ${}^3\Pi_u$ state. Hence, there
exists quite large domain of magnetic fields where the
$\rm{H}_2$-molecule is unbound being represented by two hydrogen
atoms in the same electron spin state but situated at infinite
distance from each other. The optimal geometry of the
$\rm{H}_2$-molecule (when exists) corresponds always to the elongation
along a magnetic line for the ${}^1\Sigma_g$ state thus forming a
finite chain. It is assumed that the chain in the ${}^3\Pi_u$ state
is stable towards the deviation from linearity. This assumption
seems well justified from physics point of view at large magnetic
fields: any deviation from linearity leads to a sharp increase in
total energy due to non-vanishing rotational energy. This chain is
stable (when exists) for all studied magnetic fields. However, this
chain always has the total energy higher than the $\rm{H}_3^+$ chain
(see below) and thus less preferable energetically. Calculations for the ${}^3\Pi_u$ state of ${\rm H}_2$ using a single function of the form (\ref{psi}) for which all $\alpha$ parameters are different (general term) are presented in Tables ~\ref{2e_E}-\ref{2e_R}.

\item[$\bf{n=3.}$]

In \cite{Turbiner:2007_H3} it is shown that the $\rm{H}_3^+$
molecular ion exists in a magnetic field as a bound state. For
$B \gtrsim 0.2$\,a.u. the ground state geometry is realized in the
linear, parallel to the magnetic field line configuration. Thus, the
three-proton finite chain occurs. In the domain $0.2 \lesssim B
\lesssim 20$\,a.u. the ground state is realized by ${}^3\Si_u$ state and
it is weakly bound. However, at $B > 20$\,a.u. the ground state ${}^3\Pi_u$ state is strongly bound and the chain is stable.

\item[$\bf{n=4.}$]

In field-free case the system $(4p 2e)$ does not display any
binding. However, for magnetic fields $B \gtrsim 2000$ \,a.u. it
becomes bound in the linear configuration aligned along the magnetic
line with the ${}^3\Pi_u$ state as the ground state. Hence, the
molecular ion $\rm{H}_4^{2+}$ begins to exist. Its total energy is
lower systematically than the total energy of $\rm{H}_3^{+}$. Hence,
the molecular ion $\rm{H}_4^{2+}$ is stable. With an increase of the
magnetic field strength, the total energy at the equilibrium
position decreases, the system becomes more bound (in this case, the
double ionization energy is $E_I=-E_T$, it increases with $B$) and
more compact (the internuclear equilibrium distance decreases with
$B$). Eventually, we state that the finite chain $\rm{H}_4^{2+}$
is always stable. For magnetic fields
$1 \lesssim B \lesssim 2000$ \,a.u. the state ${}^3\Pi_u$ is bound but the
ground state corresponds to an unbound system in the repulsive
${}^3\Sigma_u$ state: It consists of two $\rm{H}_2^+$ ions at
infinite distance from each other.

\item[$\bf{n=5.}$]

In field-free case the system $(5p 2e)$ does not display any
binding. However, for magnetic fields $B \gtrsim 5000$ \,a.u. it
becomes bound in the linear configuration aligned along the magnetic
line with the ${}^3\Pi_u$ state as the ground state;  hence, the
molecular ion $\rm{H}_5^{3+}$ begins to exist. For $5000 \lesssim B \lesssim 10^6$\,a.u. the ${\rm H}_5^{3+}$ molecular ion decays to ${\rm H}_4^{2+} + {\rm p}$. At magnetic fields $B \gtrsim 10^6$\,a.u. the molecular ion ${\rm H}_5^{3+}$ becomes stable.

\item[$\bf{n=6.}$]

It is not found an indication to the bound state of the $(6p2e)$-system even for the highest magnetic field studied.

\end{enumerate}

\subsection{\label{hchains2eC}Results}

The results of the calculations are presented in Tables ~\ref{2e_E}-\ref{2e_R}.
Three traditional for the field-free case systems $\rm{H}^-$, $\rm{H}_2$ and $\rm{H}_3^{+}$ continue to exist at magnetic fields $10^2 \rm{a.u.}
\leq B \leq 10^7$\,a.u. The first exotic molecular system $\rm{H}_4^{2+}$
appears at $\sim 2 \times 10^3$\,a.u. in linear configuration and
exists for all larger magnetic fields. Another exotic molecular system
$\rm{H}_5^{3+}$ appears at slightly larger magnetic field
$\sim 5 \times 10^3$\,a.u. No other two-electron molecular hydrogenic
system is seen in the domain $B \leq 10^7$\,a.u. At large magnetic fields
the ground state of all studied systems is the spin-triplet state with
spin projection $m_s=-1$ and total magnetic quantum number $m=-1$
in agreement with Ruderman hypothesis.
For $n>1$ the optimal geometry of the molecular system is linear, and
the system is aligned along magnetic field. Thus, each molecular system
forms a finite chain. It is checked that such a linear configuration
is stable with respect
to small vibrations and its vibrational energies can be calculated.
However, we were not able to check stability of the configuration
with respect to small deviations from linearity and to calculate the
rotational energies. All studied finite chains are characterized by
two features: with a magnetic field growth (i) their binding
energies increase and (ii) their longitudinal lengths decrease -
each system becomes more bound and compact. For all studied magnetic
fields $B \gtrsim 10^2$\,a.u. the systems $\rm{H}^-$ and $\rm{H}_2$ are
stable. They are characterized by much smaller binding energies in
comparison with other systems. Thus, their significance for a
thermodynamics at a fixed magnetic field seems limited.

It is worth emphasizing that among two-electron hydrogenic finite chains the
system H$^+_3$ has the lowest total energy in the domain $10^2 \lesssim B
\lesssim 2 \times 10^3$\,a.u., at larger magnetic fields $2 \times 10^3
\lesssim B \lesssim 10^6$\,a.u. the finite chain H$^{2+}_4$ gets the lowest
total energy and, eventually, at $B \gtrsim 10^6$\,a.u. the molecular ion
${\rm H}_5^{3+}$ (the longest hydrogenic chain) is characterized by the
lowest total energy. Interestingly, in the domain $10^6 \lesssim  B \lesssim 10^7$\,a.u. {\it all} two-electron finite Hydrogen chains are stable.

\section{\label{hechains1e}One-electron helium and helium-hydrogen chains}

\subsection{\label{hechains1eA}Generalities}

Let us consider now molecular systems composed of one electron and a
finite number $n$ of infinitely-massive protons and/or
$\alpha$-particles as charged centers, situated on a line which
coincides to the direction of an homogeneous magnetic field. The geometrical
arrangement is similar to that depicted in Figure~\ref{figure1}, except for the fact that charged centers can be either protons or $\alpha$ particles.
If found, bound states of such
systems are called one-electron helium or helium-hydrogen chains.  In the
present review only one-electron helium or helium-hydrogen chains
with $n=1,2,3$ were included.

Following similar considerations as for the case of hydrogenic
chains (see section~\ref{hchains1e}), the Hamiltonian which describes the one-electron helium
(helium-hydrogen) chains in a magnetic field  oriented along the $z$
direction, ${\bf B}=(0,0,B)$ is given by the Hamiltonian
(\ref{H1e-final}) with $Z_i,Z_j=1$ or $2$, depending on each
particular system. Since we are interested by the ground state for
which $m=0$ and $m_s=-1/2$, the last term in (\ref{H1e-final}) can
be omitted and the reference point for energy becomes equal to $(-B)$.

\subsection{\label{hechains1eB}Method}
The variational method is used for a study of the helium
(helium-hydrogen) chains described by the Hamiltonian
(\ref{H1e-final}). Trial functions are chosen following physics
relevance arguments \cite{turbinervar}. Their explicit expressions are
linear superpositions of $K$ terms given by  functions of the class
(\ref{1e-psi}), where $A_k$ and $\al_{k,i}, \be_k$ are linear and
non-linear parameters, respectively. Internuclear distances $R$ are
considered as variational parameters as well. In this case the
notation $\{  \}$ in (\ref{1e-psi}) means the symmetrization of the
expression inside the brackets with respect to the permutations of
the identical charged centers (for example for the system ${\rm
(HHeH)}_{}^{3+}$ it means permutation with respect to the external
protons. As for the case of hydrogenic chains, each term in
(\ref{1e-psi}) has a certain physical meaning (see section \ref{hchains1e}).  In
the following we describe the different chains
studied.

\begin{enumerate}
\item[$\bf{n=1.}$]
$(\alpha e)$.
This case corresponds to the simplest one electron helium system. It
is known that the positive  atomic ion of helium exists for any  magnetic field strength.
Furthermore, it is the only one electron helium system which
exists for magnetic fields of strength $B\lesssim 10\,$a.u.
The results presented below
for the ground state $1s_0$ of the ${\rm He}^+$ atomic ion (see Table~\ref{1e_E-He})
were obtained with a seven-parametric variational trial function introduced in
\cite{turbiner-pot:2001} for a study of the ${\rm H}$-atom.

\item[$\bf{n=2.}$]

\begin{enumerate}
  \item[(i)] $(\alpha\alpha e)$.
      Accurate variational calculations  in equilibrium  configuration (parallel to the magnetic field) for the ground state $1\sigma_g$ of the system ${\rm He}_{2}^{3+}$ were carried out in details in \cite{Turbiner:2006,turbiner2007} for the range of magnetic fields $10^2\,{\rm a.u.}\lesssim B \lesssim B_{\rm Schwinger}$. A 3-term trial function of the form (\ref{1e-psi}) which depends on ten free parameters including the internuclear distance $R$ is used in the calculations. It is the same  linear superposition of the Heitler-London,  Hund-Mulliken and Guillemin-Zener wavefunctions which was used to study the ${\rm H}_2^+$ molecular ion (see section \ref{hchains1e} above). It is found that for magnetic fields $10^2 \lesssim B \lesssim 10^3\,$a.u. the system  ${\rm He}_{2}^{3+}$ is  unstable towards the decay to ${\rm He}^+ + \alpha$. Nonetheless, at $B\gtrsim 10^4\,$a.u. this compound becomes the system with the lowest total energy among the one electron helium (helium-hydrogen) chains. In \cite{turbiner2007} lowest vibrational and rotational energies for this system were also calculated.

  \item[(ii)] $(\alpha p e)$. The first indication about the existence of the hybrid system ${\rm (HeH)}_{}^{2+}$,   for magnetic fields $B\gtrsim 10^{4}\,$a.u., was  established in \cite{Turbiner:2006,turbiner2007}, where accurate variational calculations   for the ground state $1\sigma$ of the system  ${\rm (HeH)}^{2+}$ were carried out. Variational calculations are done with a 3-term trial function of the type (\ref{1e-psi}). In \cite{Turbiner:2006,turbiner2007} it was also demonstrated that the equilibrium configuration corresponds to the situation when the molecular axis (the line connecting the proton and the $\alpha$ particle) is parallel to the magnetic field. For the narrow range of magnetic fields $10^{4}\,{\rm a.u.} \lesssim B\lesssim B_{Schwinger}$ the system ${\rm (HeH)}_{}^{2+}$ is found to be a long-living metastable state decaying to ${\rm He}^+ + p$. For magnetic fields larger than $B_{Schwinger}$ the system  becomes stable towards the decay to ${\rm He}^+ + p$.

\end{enumerate}

\item[$\bf{n=3.}$]

\begin{enumerate}

   \item[(i)] $(\alpha\alpha\alpha e)$.
      It seems it is for the first time we see an indication to the possible existence of the exotic molecular ion ${\rm He}_{3}^{5+}$   for magnetic fields $B\gtrsim 10^{6}\,$a.u. For this system a 3-term trial function of the form (\ref{1e-psi}) is used for its variational study. It  depends on  22 free parameters including two internuclear distances $R_{1,2}$. This function
      is the same linear superposition of the Heitler-London,
      Hund-Mulliken and a type of the Guillemin-Zener wavefunctions which was used to study the ${\rm H}_3^{2+}$ molecular ion (see section \ref{hchains1e} above). It is found that the system $(\alpha\alpha\alpha e)$   begins to
      exist as a bound state (i.e. displays a minimum in the corresponding  potential energy surface for finite internuclear distances) at magnetic fields $B\gtrsim 10^{6}\,$a.u. in the linear symmetric configuration (for which the two internuclear distances are equal, $R_1=R_2$) parallel to the magnetic field direction. Ground state is $1\sigma_g$.

   \item[(ii)] $(p \al p e)$.
      First indications on the existence  of the exotic trilinear
      molecular ion ${\rm (H-He-H)}_{}^{3+}$ for magnetic fields $B\gtrsim B_{\rm Schwinger}$, were given in~\cite{Turbiner:2006,turbiner2007}. For this system a 3-term trial function of the form (\ref{1e-psi}) which depends on 14 free parameters including two $R_{1,2}$ is used in the variational calculations. The results clearly show the appearance of a minimum in the potential energy surface of the $(\alpha p p e)$ system for the symmetric configuration of the charged centers $(p-\alpha-p)$ with $R_1=R_2$.  Ground state is the type $1\sigma_g$. It was not seen an indication to the existence of non-symmetric configuration $(\alpha-p-p)$.

   \item[(iii)] $(\al p \al e)$.
      First indications on the existence  of the exotic trilinear
      molecular ion ${\rm (He-H-He)}_{}^{4+}$ for magnetic fields $B\gtrsim B_{\rm Schwinger}$ were given in~\cite{Turbiner:2006,turbiner2007}. For this system a 3-term trial function (\ref{1e-psi}) which depends on 14 free parameters including two internuclear distances $R_1,R_2$ is used in the variational calculations. The results show the appearance of a minimum in the potential energy surface of the $(\alpha \alpha p e)$ system for the symmetric configuration of the charged centers $(\alpha-p-\alpha)$ with $R_1=R_2$.   Ground state is $1\si_g$. It was not seen an indication to the existence of non-symmetric configuration $(\al-\al-p)$.
\end{enumerate}

\item[$\bf{n=4.}$]

      No binding is detected for systems $(\al \al \al \al e)$, $(\al p p \al e)$, $(p \al \al p e)$ even for the highest studied  magnetic field $\sim 10^7$a.u.

\end{enumerate}

\subsection{\label{hechains1eC}Results}

The results of the ground state calculations are presented
in Tables \ref{1e_E-He}-\ref{1e_R-He}. The positive atomic ion of helium He$^+$ is the only system which exists for all studied magnetic fields $0 \leq B \leq 10^7$\,a.u.
At $B \sim 10^2$\,a.u. the first exotic molecular system $\rm{He}_2^{3+}$ appears
being unstable towards decay to $\rm{He}^{+} + \alpha$ in the range of magnetic fields $10^2\,{\rm a.u.}\lesssim B \lesssim 2\times 10^4$\,a.u. For larger magnetic fields
$B \gtrsim 2\times 10^4$\,a.u. the system $\rm{He}_2^{3+}$  becomes the most bound one-electron system among the systems made out from protons and/or $\alpha$-particles and it is stable.
Two exotic molecular systems  begin to exist at about the same magnetic field $B \sim 10^4$\,a.u. Namely, the hybrid molecular ion ${\rm (HeH)}^{2+}$, followed by the trilinear symmetric molecular system ${\rm (H-He-H)}_{}^{3+}$,
being unstable towards decay to $\rm{He}^{+} + p$ and $\rm{He}^{+} + 2p$, respectively.
Remarkably, the system ${\rm (HeH)}^{2+}$ rapidly becomes stable for magnetic
fields $B \gtrsim 2\times 10^4$\,a.u. The system ${\rm (H-He-H)}_{}^{3+}$
becomes more bound than $\rm{He}^{+}$ for magnetic fields $B\gtrsim 5\times 10^5\,$a.u.
but remains unstable towards decay to ${\rm (HeH)}^{2+}+p$ in the range of magnetic fields $ 5\times 10^5\,\lesssim B \leq 10^7$\,a.u. It never dissociates to ${\rm H}_{2}^{+} + \alpha$. Another exotic symmetric molecular system ${\rm (He-H-He)}_{}^{4+}$ appears at $B\sim B_{\rm Schwinger}$, being  unstable towards decay to ${\rm (HeH)}^{2+}+\alpha$ for  magnetic fields $ B_{\rm Schwinger}\,\lesssim B \leq 6.5\times 10^6$\,a.u., as well as  towards decay to $\rm{He}_2^{3+} + p$ for all magnetic fields studied. It is worth noting that, in spite of the  greater Coulomb repulsion, the system ${\rm (He-H-He)}_{}^{4+}$ becomes more bound than ${\rm (H-He-H)}_{}^{3+}$ for  magnetic fields $B\gtrsim 1.8\times 10^6\,$a.u. The last exotic molecular system ${\rm He}_{3}^{5+}$ appears at $B\gtrsim 10^{6}$\,a.u. This system is unstable with respect to decay into $\rm{He}_2^{3+} + \alpha$. Present level of available computational resources does allow to draw a reliable conclusion about
this molecular system at larger magnetic fields.
No more one-electron helium-hydrogenic system is seen for the range of magnetic fields studied $B \leq 10^7$\,a.u.

Concrete variational calculations for the chains $\rm{He}_2^{3+}$ and ${\rm (HeH)}^{2+}$ demonstrate that
the optimal geometry of the molecular systems is linear and aligned along magnetic field, being stable with respect to small deviations from linearity. This is understood with simple arguments since any slight deviation from the magnetic field direction leads to a large increase in the rotational energy.
So, it is natural to assume that all other studied linear chains are also stable with respect to small deviations from linearity.

All studied finite chains are characterized by two features: with a
magnetic field growth (i) their total energies increase and (ii)
their equilibrium size decreases - each system becomes more bound and compact.

Summarizing, one can state that among the one-electron helium-hydrogen
chains there are two helium systems characterized by the lowest total energy for different magnetic fields: it is the $\rm{He}^+$ ion at $0 \lesssim B \lesssim  2\times 10^3$\,a.u.
and  the $\rm{He}_2^{3+}$-chain at $2 \times 10^3 \lesssim B \lesssim 10^7$\,a.u.

%%%%%%%%%%%%%%%%%%%%  TABLE V %%%%%%%%%%%%%%%%%%%%%
\begin{table*}[ht]
 \begin{center}
\small
\begin{tabular}{|c||ccccccc|}  \hline
 &  &  &  &  &  &  &              \\[-7pt]
\raisebox{-5pt}{\small System}
\raisebox{5pt}{\hspace{-15pt}$B$\,(a.u.)} & $1$ & $10$ &
$10^{2}$ & $10^{4}$ & $4.414\times 10^{13}\,$G &$10^{6}$ &\ $10^{7}$     \\[7pt]
\hline\hline
                           &   &   &   &   &   &   &  \\[-5pt]
 ${\rm He}_{}^{+}$         &\ 4.8820 &\ 8.7801 &\ 19.109 &\ 78.426 & 92.528 & 226.66 &\ 345.17\ \\[5pt]
 ${\rm He}_{2}^{3+}$     &--&--&\ 16.516 &\ 86.233 & 105.121 & 305.11 &\ 507.31\ \\[5pt]
 ${\rm He}_{3}^{5+}$     &--&--& -- & -- & -- & 227.83 &\ 417.15\ \\[5pt]
 ${\rm (HeH)}_{}^{2+}$   &--&--& -- & 77.303 & 92.858 &  251.32 &\ 402.10\ \\[5pt]
 ${\rm (HHeH)}_{}^{3+}$  &--&--& -- & 64.747 & 79.69 & 233.71 &\ 392.47\ \\[5pt]
 ${\rm (HeHHe)}_{}^{4+}$ &--&--& -- & --     & 70.76 & 230.38 &\ 408.58\ \\[5pt]
\hline
\end{tabular}
\end{center}
\caption{\label{1e_E-He} Binding energies (in Ry) for the ground
state $1\sigma_g$ of the one-electron helium and helium-hydrogenic
linear systems (finite chains) in a magnetic field (the ground state for ${\rm (HeH)}_{}^{2+}$ is $1\si$). For ${\rm He}_3^{5+}$: $E_b=86.76\,$Ry, $R_{eq}=0.202\,$a.u. at
$B=10^{14}\,$G, while for $B_{\rm Schwinger}$ there is no minimum.}
\end{table*}

 %%%%%%%%%%%%%%%%%%%%  TABLE VI %%%%%%%%%%%%%%%%%%%%%
\begin{table*}[ht]
%\begin{sidewaystable}
\small
 \begin{center}
% use packages: array
\begin{tabular}{|c||c|c|c|c|c|}  \hline
 &  &  &  &  &              \\[-7pt]
\raisebox{-5pt}{\small System} \raisebox{5pt}{$B$\,(a.u.)} &
$10^{2}$ & $10^{4}$
&$4.414\times 10^{13}\,$G & $10^{6}$ & $10^{7}$ \\[5pt]       \hline\hline
                                &   &   &   &   &             \\[-5pt]
 ${\rm He}_{2}^{3+}\quad (R)$   &\ 0.779\ &\ 0.150 & 0.126 & 0.049 & 0.032  \\[5pt]
 ${\rm He}_{3}^{5+}\quad (R,R)$ & -- & -- & -- &\ 0.070, -\ &\ 0.041, -\ \\[5pt]
 ${\rm (HeH)}^{2+}\quad (R)$    & -- & 0.142 &  0.119 & 0.048 & 0.032   \\[5pt]
 ${\rm (HHeH)}_{}^{3+}\quad (R,R)$  & -- &\ 0.227, -\ &\ 0.184, -\ &\ 0.058, -\ &\  0.035, -\  \\[5pt]
 ${\rm (HeHHe)}_{}^{4+}\quad (R,R)$ & -- & -- &\ 0.170, -\ &\ 0.051, -\ &\ 0.031, -\  \\[5pt]
 \hline
\end{tabular}
\end{center}
\caption{\label{1e_R-He}
 Internuclear equilibrium distances (in a.u.) for the ground state $1\sigma_g$ of the
one-electron helium and helium-hydrogenic linear systems (finite
chains) in a strong magnetic field (the ground state for ${\rm (HeH)}_{}^{2+}$ is $1\si$). For all configurations which have center of symmetry, symmetric internuclear distances are not displayed.}
%\bf Note: The values for the equilibrium distance in ${\rm (HeHHe)}_{}^{4+}$ and
%${\rm (HHeH)}_{}^{3+}$ at $B=10^6, 10^7\,$a.u. indicate that the minimizations were
%not finished.}
\end{table*}
%\end{sidewaystable}

\section{\label{hechains2e}Two-electron helium and helium-hydrogen chains}

\subsection{\label{hechains2eA}Generalities}

Let us consider systems of two electrons and $n$ infinitely-massive
$\alpha$-particles situated on a line which coincides to the
magnetic line. If a bound state is found the system is called
$2e$-linear Helium chain of the length $n$ indicating the existence of
the ion $\rm{He}_n^{(2n-2)+}$ in linear geometry.

The Hamiltonian which describes systems of two electrons and a number of
$\alpha$ particles with a magnetic field oriented along the $z$
direction, ${\bf B}=(0,0,B)$ is given by the Hamiltonian (\ref
{H2e-final}) with $Z_i=Z_j=2$.

All performed calculations show that in the optimal geometry the
chain possesses a symmetry property similar to two-electron
hydrogenic chains: for any $\alpha$-particle there is a partner
situated symmetrically with respect to the center of the chain.

Another type of systems we study are mixed ones: out of $n$ heavy centers some of them have the charge two ($\al$-particles) and some have the charge one (protons). If a bound state is found the system is called
$2e$-linear Helium-Hydrogen chain of the length $n$.

\subsection{\label{hechains2eB}Method}

For these systems we follow similar consideration as for the case of
two-electron hydrogenic chains. The variational procedure is used to
explore the problem. Physical relevance arguments are followed to
choose the trial function (see, e.g. \cite{turbinervar}) which is
given by the function (\ref{psi}).

%%%%%%%%%%%%%%%%%%%%  TABLE VII %%%%%%%%%%%%%%%%%%%%%
\begin{table*}[ht]
 \begin{center}
\small
\begin{tabular}{|c||ccccccc|}  \hline
 &  &  &  &  &  &  &\\[-7pt]
\raisebox{-5pt}{\small System}
\raisebox{5pt}{\hspace{-15pt}$B$\,(a.u.)}
 & $10^{2}$ & $10^{3}$ & $10^{4}$ & $4.414\times 10^{13}$\,G &$10^{5}$ & $10^{6}$
 & $10^{7}$ \\[7pt]
\hline\hline
 &  &  &  &  &  & & \\[-5pt]
${\rm He}_{}^{}$    &\ 25.65\ & 54.37 & 106.4  & 126.0  & 191.4  & 319.7  &\ 494.3        \\[5pt]
${\rm He}_{2}^{2+}$ &\ 33.98\ & 80.49 & 174.51 & 212.14 & 343.47 & 616.68 &\ 1016.75  \\[5pt]
${\rm He}_{3}^{4+}$ &\ 26.58\ & 68.93 & 163.90 & 202.60 & 352.50 & 684.19 &\ 1212.40  \\[5pt]
${\rm He}_{4}^{6+}$ &   --    &  --   &   --   &  --    & 272.07 & 576.85 &\ 1089.89
\\[5pt]
${\rm HeH}^{+}$ & 28.36 & 64.24 & 133.49 & 160.50 & 253.22 & 440.24 & 709.65
\\[5pt]
${\rm (H-He-H)}^{2+}$ & -- & -- & 142.40 & 172.58 & 279.39 & 509.99 & 843.38
\\[5pt]
${\rm (He-H-He)}^{3+}$ & -- & -- & 153.62 & 190.22 & 320.63 & 603.91 & 1029.95
\\[5pt]
${\rm (H-He-He-H)}^{4+}$  &  &  &  &  & 275. & 585.0 & 979.1
\\[5pt]
${\rm (He-H-H-He)}^{4+}$  &  &  &  &  & 223. & 510.4 & 885.2
\\[5pt]
\hline
\end{tabular}
\end{center}
\caption{\label{2e_E-He} Double ionization energies $E_I$ in Ry
for the ground state ${}^3\Pi_u$ of the two-electron
helium and helium-hydrogenic linear systems (finite chains) in a strong
magnetic field (the ground state for ${\rm (HeH)}_{}^{+}$ is ${}^3\Pi$). }
\end{table*}

%%%%%%%%%%%%%%%%%%%%  TABLE VIII %%%%%%%%%%%%%%%%%%%%%
%\begin{sidewaystable}
\begin{table*}[ht]
\small
 \begin{center}
% use packages: array
 \begin{tabular}{|c||c|c|c|c|c|c|c|}  \hline
 &  &  &  &  &  &  &\\[-7pt]
\raisebox{-5pt}{\small System}
\raisebox{5pt}{\hspace{-15pt}$B$\,(a.u.)}
 & $10^{2}$ & $10^{3}$ & $10^{4}$ & $4.414\times 10^{13}$\,G &$10^{5}$ & $10^{6}$
 & $10^{7}$ \\[7pt]
\hline\hline
 &  &  &  &  &  & & \\[-5pt]
${\rm He}_{2}^{2+}\, (R) $ & 0.463  &  0.212  & 0.106   &  0.0902  & 0.060 & 0.0353 & 0.023  \\[5pt]
${\rm He}_{3}^{4+}\, (R,R)$ & 0.67, - & 0.27, - & 0.122, - & 0.116, - & 0.063, - & 0.0358, -  & 0.023, - \\[5pt]
${\rm He}_{4}^{6+}\, (R_1,R_2,R_1)$ & -- & --  & --  & --  &   0.089,0.060, - &  0.047, 0.037, - &  0.030,0.023, - \\[5pt]
${\rm HeH}^{+}$ & 0.440 & 0.203 & 0.104 &0.092 & 0.0585 & 0.0356 & 0.0238 \\[5pt]
${\rm (H-He-H)}^{2+}\, (R_1,R_1)$ & -- & --  & 0.105, - & 0.092, - & 0.059, - & 0.035, - & 0.022, - \\[5pt]
${\rm (He-H-He)}^{3+}\, (R_1,R_1)$ & -- & --  & 0.095, - & 0.081, - & 0.051, - & 0.030, - & 0.018, - \\[5pt]
${\rm (H-He-He-H)}^{4+}$     &  &  &  &  &  & &     \\[-5pt]
$\scriptsize (R_1,R_2,R_1)$  &  &  &  &  & 0.07, 0.10, - &  0.047, 0.030, -& 0.027, 0.015, - \\[-5pt]
${\rm (He-H-H-He)}^{4+}$  &  &  &  &  &  & & \\[-5pt]
$\scriptsize (R_1,R_2,R_1)$  &  &  &  &  & 0.08, 0.12,- & 0.041, 0.025, -&  0.025, 0.019, -\\
\hline
\end{tabular}
\end{center}
\caption{\label{2e_R-He} Internuclear equilibrium distances (in
a.u.) for the ground state ${}^3\Pi_u$ of the two-electron helium
linear systems (finite chains) in a strong magnetic field (the ground state for ${\rm (HeH)}_{}^{+}$ is the ${}^3\Pi$ state). All configurations (except for ${\rm (HeH)}_{}^{+}$) have center of symmetry and symmetric interproton distances are not displayed.}
\end{table*}
%\end{sidewaystable}
\begin{enumerate}

\item[{$\bf{n=1.}$}] $(\al e e)$

This case is only mentioned for the sake of completeness. It is
known that the helium atom exists for any magnetic field strength
\cite{schmelcherHe}. At zero field and as well as for small magnetic
fields $B \lesssim 0.75$\ a.u. the spin-singlet state $1 {}^1 0^+$ is
the ground state. For $B \gtrsim 0.75$ a.u. the spin-triplet state
$1 {}^3(-1)^{+}$ becomes the ground state. Neutral Helium atom is the least
bound system among two-electron Coulomb systems made from $\alpha$-particles.

\item[$\bf{n=2.}$]

\begin{enumerate}

   \item[(i)] $(\al \al e e)$

     The $\rm{He}_2^{2+}$ -molecule was studied in details in
     \cite{Turbiner:2006_he2ee} in a magnetic field $B=0-4.414 \times 10^{13}\,$\,G. It was shown that the lowest total energy state depends on the magnetic field strength. Similarly to the case of $p p e e$, it evolves from the spin-singlet ${}^1\Si_g$ metastable state at $0 \leq B \lesssim 0.85$\,a.u. to a repulsive spin-triplet ${}^3\Si_u$ state (unbound state) for $0.85\,\mbox{a.u.} \lesssim \ B \lesssim 1100$\,a.u. and, finally, to a strongly bound spin-triplet ${}^3\Pi_u$ state. Hence, there exists quite large domain of magnetic fields where the $\rm{He}_2^{2+}$-molecule is unbound being represented by two atomic helium ions in the same electron spin state but situated at the infinite distance from each other. The optimal geometry of the $\rm{He}_2^{2+}$-molecule (when exists) corresponds always to the elongation along a magnetic line forming a finite chain. It is assumed that the chain in the ${}^3\Pi_u$ state is stable towards the deviation from linearity. This chain is stable (or metastable) for all studied magnetic fields. However, this chain has the total energy higher than the ${\rm He}_3^{4+}$ chain (see below) for $B \gtrsim 3 \times 10^4$\,a.u. and thus less preferable energetically.

%\pagebreak
    \item[(ii)]  $(\al p e e)$

    It is the simplest $2e$ mixed helium-hydrogen system. A detailed study of the low-lying electronic states ${}^1\Si,{}^3\Si,{}^3\Pi,{}^3\De$ of the ${\rm HeH}^+$ ion was carried out in \cite{Turbiner:2007_hehee}. The ground state evolves from the spin-singlet ${}^1\Si$ state for small magnetic fields $B\lesssim 0.5 $\,a.u. to the spin-triplet ${}^3\Si$ (unbound or weakly bound) state for intermediate fields and to the spin-triplet strongly bound $^3\Pi$ state for $B \gtrsim 15 $\,a.u. When the $\rm{HeH}^+$ molecular ion exists, it is stable with respect to a dissociation. In the domain $B \gtrsim 15 $\,a.u. the optimal geometry is linear and parallel: the ion is elongated along a magnetic line. Hence, the chain is formed. With a magnetic field increase the chain gets more bound and more compact.  At magnetic fields $B < 10^4$\,a.u. the double ionization energy $E_I$ of the $\rm{HeH}^+$ ion is smaller but comparable with one of the $\rm{He}_2^{2+}$ ion. However, for $B > 10^4$\,a.u. $E_I$ gets, in fact, the smallest value among $2e$ Helium-contained molecular ions.

\end{enumerate}

\item[$\bf{n=3.}$]

\begin{enumerate}

   \item[(i)] $(\al \al \al e e)$

    In field-free case the system $(\al \al \al e e)$ does not display any binding. However, for magnetic fields $B \gtrsim 100$ \,a.u. the ${\rm He}_3^{4+}$-molecule becomes bound in the linear configuration aligned along the magnetic line. For $100\,\mbox{a.u.} \lesssim B \lesssim 1000$\,a.u. the ${}^3\Si_u$ state is the ground state \cite{helium3:unpublished}. This state is a metastable state for any magnetic field, its total energy lies above the total energies of its lowest dissociation channel. For $B \gtrsim 1000$\,a.u. the state ${}^3\Pi_u$ is the ground state. For magnetic fields $1000\, \mbox{a.u.} \lesssim B \lesssim 3 \times 10^4$\,a.u.  the total energy of the dominant dissociation channel ${\rm He}_3^{4+} \to {\rm He}_2^{2+}({}^3\Pi_u) + \al$  is lower than the total energy of the ${\rm He}_3^{4+}$ ion in the ${}^3\Pi_u$ state. Thus, in this range of magnetic fields, the ion ${\rm He}_3^{4+}({}^3\Pi_u)$ is a metastable state towards the lowest channel of decay. Hence, for magnetic fields $B \gtrsim 3 \times 10^4$\,a.u. the molecular ion ${\rm He}_3^{4+}$ ion in the ${}^3\Pi_u$-state is stable. With an increase of the magnetic field strength, the total energy at the equilibrium position decreases, the system becomes more bound (in this case, the double ionization energy is $E_I=-E_T$, it increases with $B$) and more compact (the internuclear equilibrium distance decreases with $B$).

\item[(ii)] $(p \al p e e)$

    In field-free case the system $(p \al p e e)$ does not display any
    binding. However, for magnetic fields $B \gtrsim 10^4$ \,a.u. the ${\rm (H-He-H)}^{2+}$-ion becomes bound in the linear configuration aligned along the magnetic line with the ${}^3\Pi_u$ state as the ground state. This ion is stable.

\item[(iii)] $(\al p \al e e)$

    In field-free case the system $(\al p \al e e)$ does not display any
    binding. However, for magnetic fields $B \gtrsim 10^4$ \,a.u. the ${\rm (He-H-He)}^{3+}$-ion becomes bound in the linear configuration aligned along the magnetic line with the ${}^3\Pi_u$ state as the ground state. This ion is unstable towards a decay to ${\rm He}_2^{2+}({}^3\Pi_u) + p$, however, at $B > 10^6$ \,a.u. the ion ${\rm (He-H-He)}^{3+}$ becomes stable.

\end{enumerate}

\item[$\bf{n=4.}$]

\begin{enumerate}

   \item[(i)] $(4\al 2e)$

    In field-free case the system $(4\al 2e)$ does not display any
    binding. However, for magnetic fields $B \gtrsim 10^5$ \,a.u. the ${\rm He}_4^{6+}$-molecule becomes bound in the linear configuration aligned along the magnetic line with the ${}^3\Pi_u$ state as the ground state. With an increase of the magnetic field strength, the total energy at the equilibrium position decreases, the system becomes more bound (in this case, the double ionization energy is $E_I=-E_T$, it increases with $B$) and more compact (the internuclear equilibrium distance decreases with $B$). For magnetic fields $B \gtrsim 10^5$ \,a.u.  the total energy of the dominant dissociation channel ${\rm He}_3^{4+}({}^3\Pi_u) + \al$  is lower than the total energy of the ${\rm He}_4^{6+}({}^3\Pi_u)$ ion. Thus, the ion ${\rm He}_4^{6+}({}^3\Pi_u)$ is a metastable state toward the lowest channel of decay. It is also unstable towards decay to ${\rm He}_2^{2+}({}^3\Pi_u) + 2\al$ for magnetic fields $ 10^5\lesssim B \lesssim 2\times 10^6$ \,a.u.

   \item[(ii)] $(p\al\al p 2e)$

    In field-free case the system $(p\al\al p 2e)$ does not display any
    binding. However, for magnetic fields $B > 10^5$ \,a.u. the ${\rm (H-He-He-H)}_{}^{4+}$-molecule
    becomes bound in the linear configuration aligned along the magnetic
    line with the ${}^3\Pi_u$ state as the ground state.
    With an increase of the magnetic field strength, the system becomes more bound (the double ionization energy increases with $B$) and
    more compact, i.e. both, the internuclear equilibrium distance $R_1$ between a proton and the closest $\alpha$ particle, and the distance $R_2$ between the two $\alpha$ particles,  decrease with $B$. For magnetic fields $B \gtrsim 10^5$ \,a.u. the total energy of the dominant dissociation channel ${\rm He}_2^{2+}({}^3\Pi_u) + 2p$  is lower than the total energy of the ${\rm (H-He-He-H)}_{}^{4+}$ ion. Thus, the ion ${\rm (H-He-He-H)}_{}^{4+}$ is a metastable state toward the lowest channel of decay.

   \item[(iii)] $(\al p p \al 2e)$

    In field-free case the system $(\al p p \al 2e)$ does not display any binding. However, for magnetic fields $B > 10^5$ \,a.u. the ${\rm (He-H-H-He)}_{}^{4+}$-molecular ion becomes bound in the linear configuration aligned along the magnetic line with the ${}^3\Pi_u$ state as the ground state. With an increase of the magnetic field strength, the system becomes more bound (the double ionization energy increases with $B$) and more compact, i.e. both, the internuclear equilibrium distance $R_1$ between a proton and the closest $\alpha$ particle, and the distance $R_2$ between the two protons,  decrease with $B$. For magnetic fields $B \gtrsim 10^5$ \,a.u. the total energy of the dominant dissociation channel ${\rm He}_2^{2+}({}^3\Pi_u) + 2p$  is lower than the total energy of the ${\rm (He-H-H-He)}_{}^{4+}$-molecular ion, thus being a metastable state toward the lowest channel of decay.

\end{enumerate}

\item[$\bf{n=5.}$]

    The results of the analysis of 5-center, 2-electron systems is shown in Table IX. It is not found an indication to binding of the proton-free systems $(5 \al 2e)$ for the whole domain of studied magnetic fields, while $(4\al p 2e)$ gets bound at $B \sim 10^7$\,a.u. being unstable decaying towards many different finite chains. The system $(3\al 2p 2e)$ is unbound although a particular configuration $(\al p \al p \al 2e)$
    displays a minimum in the potential curve. The two $\al$-contained system are bound in both symmetric configuration - $(p \al p \al p 2e)$ and $(\al p p p \al 2e)$ - while the latter one is more bound even for magnetic field $B \sim 10^6$\,a.u. This system is unstable with dominant decay mode to $(\al p \al 2e)$. One $\al$-contained system $(p p \al p p 2e)$ is bound at $\sim 10^7$\,a.u. and it is stable(!).
    It is worth noting that the system $5p 2e$ is bound for magnetic fields $B \gtrsim 10^4$\,a.u. (see Table III and a discussion on p.15).

\item[$\bf{n=6.}$]

    It is not found an indication to the bound state of any 6-center system even for the highest magnetic field studied.

\end{enumerate}

%%%%%%%%%%%%%%%%%%%%  TABLE IX %%%%%%%%%%%%%%%%%%%%%
\begin{table}
\label{Table 9}
\centering
\begin{tabular}{|c|l|cc|c}
\hline
              &          &                                       &\\[3pt]
  & &\multicolumn{2}{|c|}{$E_T=-E_I$ (Ry)}                       &\\
  \ Composition\ &\ Configuration\ &                             &\\
              &          &$B=10^6\,$a.u. & $B=10^7\,$a.u.        &\\[5pt]
\hline \hline
              &          &                                       &\\
  1-$\alpha$ 4-$p$ & H-H-He-H-H    &  Unbound   &    Bound     &\\[5pt]
                 &&  $\sim$  -450 &         -866.0 &\\[1pt]
                 &&        & {\small ($R_1=0.0228, R_2=0.0203$\,a.u.)}&\\[5pt]
\hline
              &          &                                       &\\
    2-$\alpha$ 3-$p$ & He-H-H-H-He & Bound & Bound &  \\[5pt]
                     &             & -414.5 & -792.6 &  \\[5pt]
                     & H-He-H-He-H & Bound & Bound &  \\[5pt]
                     &             & -485.3 & -873.9 &  \\[1pt]
                     &&& {\small ($R_1=0.0306, R_2=0.0189$\,a.u.)}& \\[5pt]
\hline
              &          &                                  &\\
    3-$\alpha$ 2-$p$ & He-H-He-H-He & Unbound & ``Bound"       &\\[5pt]
      &&            $\sim$  -420 &                -860.0    &\\[5pt]
      &&& {\small ($R_1=0.023, R_2=0.018$\,a.u.)}& \\[5pt]
                     & H-He-He-He-H &  Unbound & Unbound    &\\[5pt]
                        &             & $\sim$ -620 & $\sim$ -1055   &\\[5pt]
\hline
              &          &                              &\\
    4-$\alpha$ 1-$p$ & He-He-H-He-He & Unbound & Bound  &\\[5pt]
                            &             & $\sim$ -380 & -862.4 &\\[5pt]
     &&& {\small ($R_1=0.0356, R_2=0.0195$\,a.u.)}      &\\[5pt]
\hline
\end{tabular}
\caption{2-electron 5-center molecular ions (finite chains) in a magnetic field in ${}^3\Pi_u$ state -
symmetric, spin-triplet configuration parallel to the magnetic field direction. $E_T, E_I$ is total and double ionization energy, respectively.
For unbound states a characteristic total energy indicated.
}
\end{table}

\subsection{Results}
The results of the calculations are presented in Tables VII-IX. Three
traditional systems $\rm{He}$, $\rm{He}_2^{2+}$ and $\rm{HeH}^{+}$ exist for all studied magnetic fields $10^2  \leq B \leq 10^7$\,a.u. The first exotic molecular system $\rm{He}_3^{4+}$
appears at $\sim 100$\,a.u. in linear configuration and exists for larger magnetic fields. For $100\, \lesssim B \lesssim 5 \times 10^4$\,a.u. the ${\rm He}_3^{4+}$ ground state is a metastable state with respect to its lowest dissociation channel. For magnetic fields $B > 5 \times 10^4$\,a.u. the ground state of the system ${\rm He}_3^{4+}$ becomes a strongly bound state. Another exotic molecular system $\rm{He}_4^{6+}$ appears at $\sim 10^5$\,a.u. as a metastable state. No other two-electron molecular helium systems are seen for $B \leq 10^7$\,a.u. At large magnetic field the ground state of all studied systems is the spin-triplet state with spin projection $m_s=-1$ and total magnetic quantum number $m=-1$. For $n>1$ the optimal geometry of the molecular system is linear, and the system is aligned along magnetic field. Thus, each molecular system forms a finite chain. It
is checked that such a linear configuration is stable with respect
to small vibrations and its vibrational energies can be calculated.
However, we were not able to check stability of the configuration
with respect to small deviations from linearity and to calculate the
rotational energies. All studied finite chains are characterized by
two features: with a magnetic field growth (i) their binding
energies increase and (ii) their longitudinal lengths decrease -
each system becomes more bound and compact.

It is  worth noting that among two-electron helium finite chains the
system ${\rm He}_2^{2+}$ in triplet ${}^3\Pi_u$ state has the lowest
total energy in the domain $10^2 \lesssim B \lesssim 3 \times 10^4$\,a.u.,
whereas at larger magnetic fields $3 \times 10^4 \lesssim B \lesssim 10^7$\,a.u.
the finite chain ${\rm He}_3^{4+}$ in triplet ${}^3\Pi_u$ state acquires the lowest
total energy. In the domain $2\times 10^6 \lesssim  B \lesssim 10^7$\,a.u. all
 studied two electron finite helium chains become stable with the only exception
 of ${\rm He}_4^{6+}$.

\section*{\protect\bigskip \large Conclusions}

A complete non-relativistic classification of one-two electron finite molecular chains (polymers) made out of protons/$\al-$particles in a strong magnetic field is presented. It is naturally assumed that the ground state of any one-electron chain is $1\si_g$ (or $1\si$ for non-symmetric systems), while for any two-electron chain is spin-triplet ${}^3\Pi_u$ (or ${}^3\Pi$ for non-symmetric systems). All calculations were carried out in variational methods with state-of-the-art trial functions. Protons and $\al-$particles are assumed to be infinitely-massive and situated along a magnetic line.

It is clearly seen the existence of three magnetic field thresholds
\footnote{A notion of the existence of the molecule in the Born-Oppenheimer approximation is ambiguous (for a discussion see e.g. \cite{Herzberg}). In one definition it is enough for the existence if a potential curve has a minimum, in other one it is required the existence at least one vibrational, one rotational states. We follow the first definition, however, localizing a moment of the appearance of the minimum of the potential curve very approximately.},
\[
  B_t^{(1)} \sim 10^2 a.u.\ ,\ B_t^{(2)} \sim 10^4 a.u.\ ,\ B_t^{(3)} \sim 10^6 a.u.\ .
\]
At magnetic fields $B \lesssim 10^2$\,a.u. the only traditional ions, atoms and molecules may exist, the chains are not well-pronounced, they are very short containing at most two heavy particles. However, at $10^2 < B <10^4$\,a.u. several new exotic ions appear in addition to traditional ones.
All ions immediately form strongly bound linear chains aligned along a magnetic field. At $B \sim 10^4$\,a.u. several more new exotic ions appear quickly forming linear chains. Then similar appearance of new exotic ions happens at $B \sim 10^6$\,a.u. It is quite interesting that the ions which already appeared (existed) below some magnetic field threshold, above of the threshold they become stable. It is worth noting that for fixed magnetic field the neutral systems are always the least bound ones.

Concluding we present a list of 1-2$e$ proton-$\al$particle contained ions for which a certain magnetic fields exist where they are stable,
\[
  \rm{H}\ ,\ \rm{H}_2^{+}\ ,\ \rm{H}_3^{2+}\ ,
  \rm{He}^+\ ,\ \rm{He}_2^{3+}\ ,\ \rm{(HeH)}^{2+}\ ,
  \]
  \[
  \rm{H}^-\, ,\ \rm{H}_2\, ,\ \rm{H}_3^{+}\,,\ \rm{H}_4^{2+}\, ,\ \rm{H}_5^{3+}\ ,\
  \rm{He}\, ,\ \rm{He}_2^{+}\, ,\ \rm{He}_3^{4+}\, ,\  \rm{(HeH)}^{+}\, ,\
  \rm{(HHeH)}^{2+}\, ,\ \rm{(HeHHe)}^{3+}\, ,\ \rm{(HHHeHH)}^{4+}\ ,
\]
among the 25 Coulomb 1-2$e$ systems which (can) exist in a magnetic field (see Tables I-IX).

All presented results are obtained in non-relativistic way with an assumption that masses of heavy particles are infinite. They can be considered as an indication to a new atom-molecular physics in magnetic fields $B \gtrsim 10^2$\,a.u. It encourages us to an exploration of finite mass effects in a magnetic field. This issue looks quite complicated technically due to absence of a separation of variables, especially, in the case of more than two particles and non-zero total charge of the system. Those two cases are exactly ones which are the most important from the point of view of obtained results: the most bound systems contain usually more than two bodies and charged. Another important issue is related to relativistic corrections to our non-relativistic results. Although in our understanding the Duncan qualitative argument \cite{Duncan} sounds physically, it needs to be checked quantitatively. Present authors plan to study both issues in near future.

\begin{acknowledgments}

Present work took more than three years of intense dedicated studies.
The authors want to express their deep gratitude to M.I.~Eides (UK), D.~Page (IA-UNAM) and G.G.~Pavlov (PennState) for their permanent interest to the present work, regular useful discussions and the encouragement during this time.

N.L.G. is grateful to ICN-UNAM where the present study was initiated during his
Postdoc Fellowship period.

Computations were mostly performed on a dual core DELL PC with two Xeon processors of 3.06\,GHz each (ICN) and 54-node FENOMEC cluster {\it ABACO} (IIMAS, UNAM). Some test calculations were also done in the UNAM HP-CP 4000 cluster \textit{KanBalam} (Opteron).

This work was supported in part by the University program FENOMEC (UNAM),
the CONACyT grants {\bf 47899-E,\ 58942-F} (Mexico) and the PAPIIT grants
{\bf IN121106-3,\ IN115709-3} (UNAM, Mexico).
\end{acknowledgments}

%%%%%%%%%%%%%%%%%%%%%%%%%%%%%%%%%%%%%%%%%%%%%%%%%%%%
%%%%%%%%%%%%%Bibliography%%%%%%%%%%%%%%%%%%%%%%%%%%%
%%%%%%%%%%%%%%%%%%%%%%%%%%%%%%%%%%%%%%%%%%%%%%%%%%%%


\begin{thebibliography}{99}

\bibitem{Liberman:1995}
        M.A.~Liberman and B.~Johansson,
        `Properties of matter in ultrahigh magnetic fields and
         the structure of the surface of neutron stars',\\
        {\it Soviet Phys. - Usp. Fiz. Nauk. \bf 165}, 121 (1995)\\
        {\it Sov. Phys. Uspekhi \bf 38}, 117 (1995)
        (English Translation)

\bibitem{Lai:2001}
        D.~Lai, `Matter in strong magnetic fields',
        {\it Rev. Mod. Phys.\bf 73}, 629 (2001)\\
       (astro-ph/0009333)

\bibitem{Turbiner:2006}
        A.V.~Turbiner and J.C.~L\'opez Vieyra, `One-electron Molecular Systems in a Strong Magnetic Field',
        {\it Phys. Repts. \bf 424}, 309 (2006)

\bibitem{Kadomtsev:1971}
        B.B.~Kadomtsev, V.S.~Kudryavtsev,
        {\it Pis'ma ZhETF \bf 13}, 15, 61 (1971);\\
        {\it Sov. Phys. JETP Lett. \bf 13}, 9, 42 (1971)
        (English Translation)\\
        {\it ZhETF \bf 62}, 144 (1972);\\
        {\it Sov. Phys. JETP \bf 35}, 76 (1972) (English Translation)

\bibitem{Ruderman:1971}
        M.~Ruderman,  {\it Phys. Rev. Lett. \bf 27}, 1306 (1971);\\
        in IAU Symp. 53, {\it Physics of Dense Matter},
        ed. by C.J.~Hansen \\(Dordrecht: Reidel, 1974) p.117

\bibitem{Turbiner:2006London}
        A.V.~Turbiner,
        `Molecular systems in a Strong Magnetic Field -
        how atomic - molecular physics in a strong magnetic field
        might look like',\\
%           Preprint ICN-UNAM 06-03, June 2006, pp.10\\
        {\it Astrophysics and Space Science, \bf 308} , 267-277 (2007)
%         DOI 10.1007/s10509-007-9337-7

\bibitem{schmelcher-H2}
        T. Detmer, P. Schmelcher, and L. S. Cederbaum,
        {\it Phys. Rev. \bf A57}, 1767 (1998)

\bibitem{Turbiner:2007_H3}
        A.V.~Turbiner, N.L.~Guevara and J.C.~L\'opez Vieyra,\\
       `The Ion $\rm{H}_3^{+}$ in a Strong Magnetic Field.
            Linear Configuration',\\
        {\it Astrophysics and Space Science \bf 308}, 497-501 (2007) \\
%             DOI 10.1007/s10509-007-9355-5
        `The $\rm{H}_3^+$ molecular ion in a magnetic field:
         linear parallel configuration',\\
        {\it Phys.Rev. \bf A75}, 053408 (2007)\\
        (physics/0606083)

\bibitem{Duncan}
        Robert C.~Duncan, Physics in Ultra-strong Magnetic Fields,
        arXiv: astro-ph/0002442 (2000)

\bibitem{turbinervar}
        A.V.~Turbiner,
%           ``On Perturbation Theory and Variational Methods in Quantum
%           Mechanics'',\\
        {\it  ZhETF \bf 79}, 1719 (1980)\\
        {\it Soviet Phys.-JETP \bf 52}, 868 (1980)
        (English Translation);\\
        {\it Usp. Fiz. Nauk. \bf 144}, 35 (1984)\\
        {\it Sov. Phys. -- Uspekhi \bf 27}, 668 (1984)
        (English Translation);\\
        {\it Yad. Fiz. \bf 46}, 204 (1987)\\
        {\it Sov. Journ. of Nucl. Phys. \bf 46}, 125 (1987)
        (English Translation);\\
        Doctor of Sciences Thesis, ITEP, Moscow, 1989
        (unpublished),\\
        `Analytic Methods in Strong Coupling Regime (large
        perturbation) in Quantum Mechanics'

\bibitem{turbiner-pot:2001}
        A.Y.~Potekhin and A.V.~Turbiner,
        {\it Phys. Rev. \bf A63}, 065402 (2001)\\
%  `Hydrogen atom in a magnetic field: quadrupole moment',
              (physics/0101050)

\bibitem{schmelcher-H-}
        O.-A.~Al-Hujaj and  P.~Schmelcher, {\it Phys. Rev. \bf A61}, 063413 (2000)

\bibitem{Turbiner:unpublished}
        A.V.~Turbiner, J.C.~L\'opez Vieyra and N.L.~Guevara, `The $H^-$ ion in a strong magnetic field' (unpublished)

\bibitem{schmelcherHe}
        W.~Becker and P.~Schmelcher, {\it J.Phys. \bf B33}, 545 (2000)

\bibitem{Turbiner:2006_he2ee}
        A.V.~Turbiner and N.L.~Guevara,
        {\it Phys.Rev. \bf A74}, 063419 (2006)\\
        (astro-th/0610928)

\bibitem{Turbiner:2007_hehee}
        A.V.~Turbiner and N.L.~Guevara,
%       `The $\rm{HeH}^+$ molecular ion in a magnetic field'\\
%             March 2007, pp.14\\
        {\it Journ.Phys. \bf B40}, 3249-3257 (2007)\\
                (physics/0703090)

\bibitem{turbiner2007}
        A.V.~Turbiner and J.C.~L\'opez Vieyra,
        {\it Int. Jour. Mod. Phys. \bf A 22}, 1605-1626 (2007)

\bibitem{helium3:unpublished}
        A.V.~Turbiner, J.C.~L\'opez Vieyra and N.L.~Guevara,
        `The ${\rm He}_3^{4+}$, ${\rm He}_4^{6+}$ and ${\rm He}_5^{8+}$ molecular ions in a strong magnetic field' (unpublished)

\bibitem{Herzberg}
           G.~Herzberg, "Molecular Spectra and Molecular Structure. I.
           Spectra of Diatomic Molecules", Krieger
           Publishing Company, Malabar, Florida, 1989
           (Second Edition)

\end{thebibliography}
\end{document}